\documentclass[aps,prd,nopacs,nofootinbib,notitlepage,superscriptaddress,twocolumn]{revtex4-1}

\usepackage[utf8]{inputenc}
\usepackage{color}
\usepackage{amsfonts}
\usepackage{graphicx}
\usepackage[dvipsnames]{xcolor}
\definecolor{refs}{RGB}{245,156,74}
\usepackage[colorlinks=true,hyperfootnotes=true,citecolor=cyan]{hyperref}
\usepackage{tabularx,booktabs}
\newcolumntype{Y}{>{\centering\arraybackslash}X}

\usepackage{ulem}
\usepackage{dsfont} 

\pdfoutput=1

\usepackage{amsmath,amsthm,amssymb}
\usepackage{bm}

\usepackage{dcolumn}
\allowdisplaybreaks[1]

\newcommand{\newc}{\newcommand}

\newc{\rk}[1]{{\color{red} #1}}
\newc{\D}{\partial}
\newc{\rH}{{\rm H}}
\newc{\cH}{{\mathcal H}}
\newc{\dphi}{\delta\phi}
\newc{\pa}{\partial}
\newc{\tp}{\dot{\phi}}
\newc{\ttp}{\ddot{\phi}}
\newc{\drhoc}{\delta\rho_c}
\newc{\aB}{\alpha_{\rm B}}
\newc{\aK}{\alpha_{\rm K}}
\newc{\aM}{\alpha_{\rm M}}
\newc{\bn}{\beta_{n_c}}
\newc{\bK}{\beta_{\rm K}}
\newc{\delc}{\delta_{c{\rm N}}}
\newc{\eH}{\epsilon_{\rm H}}
\newc{\ep}{\epsilon_{\phi}}

\newc{\cs}{c_{\rm s}}
\newc{\dd}{{\rm de}}

\newcommand{\jcap}{J. Cosmol. Astropart. Phys.}   

\begin{document}
\title{On the (non-)degeneracy of massive neutrinos and elastic interactions in the dark sector}

\author{Jose Beltr\'an Jim\'enez}
\email{jose.beltran@usal.es}
\affiliation{Departamento~de~F{\'i}sica~Fundamental~and~IUFFyM,~Universidad~de~Salamanca,~E-37008~Salamanca,~Spain.}
\author{David Figueruelo}
\email{davidfiguer@usal.es}
\affiliation{Departamento~de~F{\'i}sica~Fundamental~and~IUFFyM,~Universidad~de~Salamanca,~E-37008~Salamanca,~Spain.}
\author{Florencia Anabella Teppa Pannia}
\email{fa.teppa.pannia@upm.es}
\affiliation{Departamento~de~Matem{\'a}tica~Aplicada~a~la~Ingenier{\'i}a~Industrial,~Universidad~Polit{\'e}cnica~de~Madrid,~E-28006~Madrid,~Spain.}

\begin{abstract}
Cosmological models featuring an elastic interaction in the dark sector have been shown to provide a promising scenario for alleviating the $\sigma_8$ tension. A natural question for these scenarios is whether there could be a degeneracy between the interaction and massive neutrinos since they suppress structures in a similar manner. In this work we investigate the presence of such a degeneracy and show that the two effects do not exhibit strong correlations.
\end{abstract}

\date{\today}
\maketitle
\section{Introduction}
\label{sec:intro}

Our current cosmological model, namely $\Lambda$CDM, has received a remarkable support from the different data that have been collected in the last decades and which include observations of the Cosmic Microwave Background~\cite{Aghanim:2018eyx}, Supernovae~\cite{SupernovaSearchTeam:1998fmf,SupernovaCosmologyProject:1998vns}, Baryon Acoustic Oscillations~\cite{SDSS:2005xqv,Ross:2014qpa,BOSS:2016wmc}, Large Scale Structures or Weak Lensing~\cite{Hildebrandt:2016iqg,DES:2017myr}, etc. As the instruments improve and the amount of data increase so does the precision on the measured cosmological parameters and some tensions between different datasets have commenced to appear~\cite{Perivolaropoulos:2021jda,DiValentino:2020vvd}. One of these tensions is the apparent discrepancy in the clustering of matter as predicted by CMB measurements~\cite{Planck:2018vyg} and those based on low redshift observations~\cite{DES:2020ahh,Planck:2015lwi}. This tension is usually described in terms of the parameter $\sigma_8$ (or the related $S_8$) that parameterises the amplitude of matter fluctuations on spheres of 8 Mpc$/h$. As usual, the tension could be driven by unknown systematics, but it could also be signalling the need for physics beyond $\Lambda$CDM. Since the tension seems to indicate that the clustering in the late-time universe appears to be smaller than what the CMB suggests, it is natural to appeal to some mechanism that erases structures or prevents the clustering at low redshift. This idea is realised in scenarios where the dark sector features some interaction between dark matter and dark energy (see Ref.~\cite{Wang:2024vmw} for a review on interactions between dark energy and dark matter). Among all the plethora of interacting models, those with a mechanism that naturally operates at late times, when precisely dark energy becomes relevant and then such a mechanism will naturally emerge, would naturally accommodate the lower clustering suggested by late probes. Moreover, if the interaction effectively provides dark matter with a pressure, that mechanism   will prevent its clustering due to such gained pressure. Furthermore, in order to leave the background cosmology unaffected, so we do not worsen the Hubble tension~\cite{DiValentino:2021izs,DiValentino:2020zio} as typically happens, one can consider the interaction to be elastic. This results in that there is only momentum exchange between the dark components at linear order. The described scenario has been investigated in several versions and in all its variants (see Refs.~\cite{Simpson:2010vh,Pourtsidou:2016ico,Asghari:2019qld,Linton:2021cgd,Chamings:2019kcl,BeltranJimenez:2021wbq,Kumar:2017bpv,Skordis:2015yra,Baldi:2016zom} for several examples where a momentum exchange takes place), being all in an agreement that the elastic interaction is efficient in alleviating the $\sigma_8$ tension. Furthermore, it has been repeatedly reported that, when including measurements of $S_8$, not only the $\sigma_8$ tension can be alleviated, but the non-interacting case is excluded at several $sigma$s and a {\it detection} of the interaction could be inferred. Also, if the interaction is indeed there, it has been shown that the dipole of the matter power spectrum might provide a smoking gun \cite{BeltranJimenez:2022irm}.

On the other hand, it is known that the presence of massive neutrinos also suppresses the growth of structures on small scales and at relatively low redshift when they become non-relativistic~\cite{Lesgourgues:2006nd,Hu:2014sea}. Thus, a natural and pertinent question to ask is whether the effects of the interaction could be degenerated with massive neutrinos since both appear to have similar effects on the matter power spectrum, i.e., they both tend to suppress the growth of structures at late times and on small scales. This degeneracy would introduce a degradation on the measured value of the interaction parameter and, hence, it would reduce the significance of the previous findings in the literature seemingly pointing towards a detection of the interaction. The goal of this work is to analyse the presence of such a degeneracy and unveil whether allowing for a varying neutrinos mass could actually degrade the measurement of the interaction parameter.

The work is structured as follows. In Sec. \ref{sec:Covariantised}
 we briefly review the model that we will consider and we will compare its effects on the matter power spectrum with those of massive neutrinos. We then will perform a MCMC analysis in Sec. \ref{sec:Results} to confront different datasets to the predictions of the interacting model while allowing for a varying neutrino mass. Finally, we will conclude in Sec. \ref{sec:Conclusions} with a discussion of our main results.


\section{The dark elastic interacting model}
\label{sec:Covariantised}
We will consider the model with an elastic interaction introduced in Ref.~\cite{Asghari:2019qld} that has been subsequently analysed in Refs.~\cite{Figueruelo:2021elm,BeltranJimenez:2021wbq,Poulin:2022sgp,Cardona:2022mdq}. The general idea is to modify the conservation equations of the dark sector by introducing an interaction that only affects the perturbations. This is achieved by exploiting the existence of a common rest frame on large scales for all the components in our universe. In this scenario, the interaction is assumed to be proportional to the relative 4-velocities of the dark fluids\footnote{The possibility of having an interaction of this type between dark energy and baryons was also explored in Ref.~\cite{BeltranJimenez:2020iyx} motivated by the study of Ref.~\cite{Vagnozzi:2019kvw}.} so the conservation equations read
\begin{eqnarray}
\nabla_\mu T^{\mu\nu}_{\rm dm} &=&\alpha(u^\nu_{\rm dm}-u^\nu_{\rm de})\,,\\
\nabla_\mu T^{\mu\nu}_{\rm de}&=&
-\alpha(u^\nu_{\rm dm}-u^\nu_{\rm de})\,,
\label{eq:coupling}
\end{eqnarray}
with $u_{\rm dm}^\mu$ and $u_{\rm de}^\mu$ the 4-velocities of cold dark matter and dark energy respectively and $\alpha$ a constant parameter that controls the strength of the interaction. As desired, this interaction will modify the standard evolution only when the relative velocity between the dark components is non-negligible. The parameter $\alpha$ has dimension $5$ in natural units and the natural scale associated to it is $\rho_c H_0$ so, from now on, we will work with a dimensionless parameter $\alpha$ that will be understood to be normalised with $\frac{3H_ 0^3}{8\pi G}$. This will be the only new parameter of this scenario and it will control both the strength of the interaction as well as the redshift at which it becomes important.

In the described scenario, the density contrast $\delta$ and velocity perturbations $\theta$ in the dark sector will be coupled not only in the usual indirect way through the gravitational potential, but they will also feature a direct coupling via the interaction. In the Newtonian gauge, the perturbation equations in the dark sector read \cite{Asghari:2019qld,Figueruelo:2021elm}:
\begin{eqnarray}
\label{eq:deltaDM}
\delta_{\rm dm}' &=& 
-\theta_{\rm dm}+3\Phi'\,,\\
\delta_{\rm de}'&=&-3 \mathcal{H}( c_{\rm de}^2-w) \delta_{\rm de} +3(1+w)\Phi'  \nonumber\\
& &-(1+w)\left(1+9 \mathcal{H}^2
\frac{c_{\rm de}^2-w}{k^2}\right)
\theta_{\rm de}\;,  \\
\label{eq:thetaDM}
\theta_{\rm dm}'&=&-\mathcal{H} \theta_{\rm dm} + k^2 \Phi + \Gamma(\theta_{\rm de}-\theta_{\rm dm})\;, \\
\label{eq:thetaDE}
\theta_{\rm de}' &=&(3c_{\rm de}^2-1) \mathcal{H}\theta_{\rm de}+k^2\Phi +\frac{k^2 c_{\rm de}^2}{1+w}\delta_{\rm de} \nonumber\\
& &-\Gamma R(\theta_{\rm de}-\theta_{\rm dm})\,,
\end{eqnarray}
where $w$ is the dark energy equation of state parameter and $c_{\rm de}^2$ its adiabatic sound speed squared. In the above equations, we have also introduced the convenient quantities $\Gamma$ and $R$ defined as
\begin{eqnarray}
\Gamma&\equiv& \alpha \frac{a^4}{\Omega_{\rm dm}} \;, \\ \label{eq:Scoupling}
R &\equiv&
\frac{\rho_{\rm dm}}{(1+w)\rho_{\rm de}}\,. \label{eq:Rcoupling}
\end{eqnarray}
These are the physically relevant quantities because $\Gamma$ measures the effective interaction rate in the dark sector, while $R$ gives the relative fraction of dark energy to dark matter. Notice that we need $w\neq-1$ to have an effect. In the strict $w=-1$ case, the dark energy component does not have perturbations and, hence, our scenario fails.\footnote{It is worth mentioning however that the limit $w\to-1$ can give non-trivial effects because this limit is not continuous and one should take it with some care.}

\begin{figure*}[!t]
\includegraphics[width=0.49\textwidth]{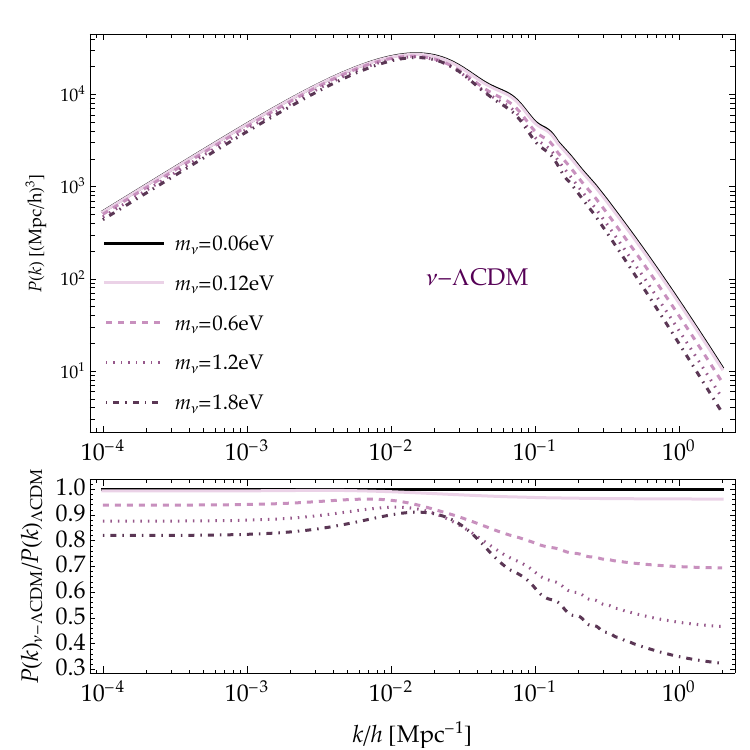}
\includegraphics[width=0.49\textwidth]{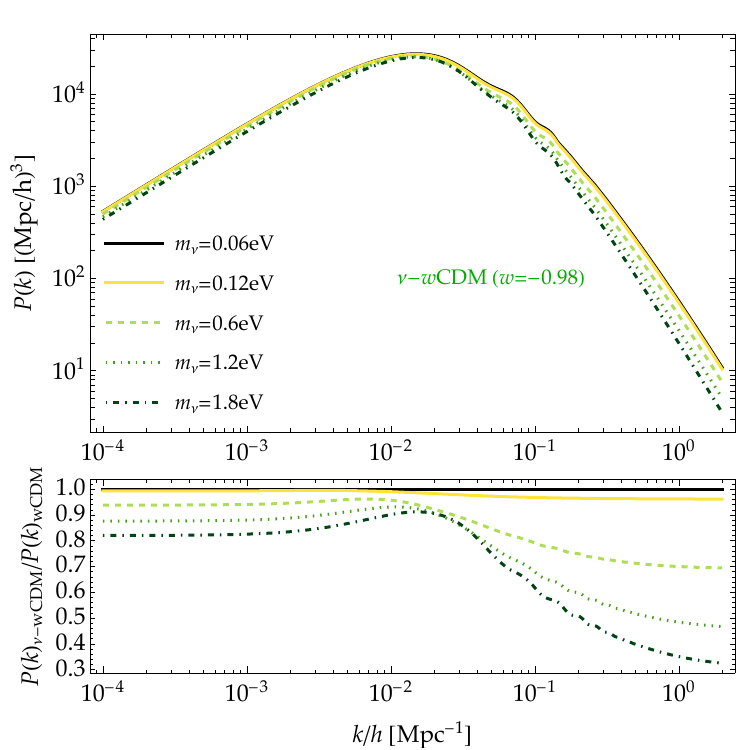} \\
\includegraphics[width=0.49\textwidth]{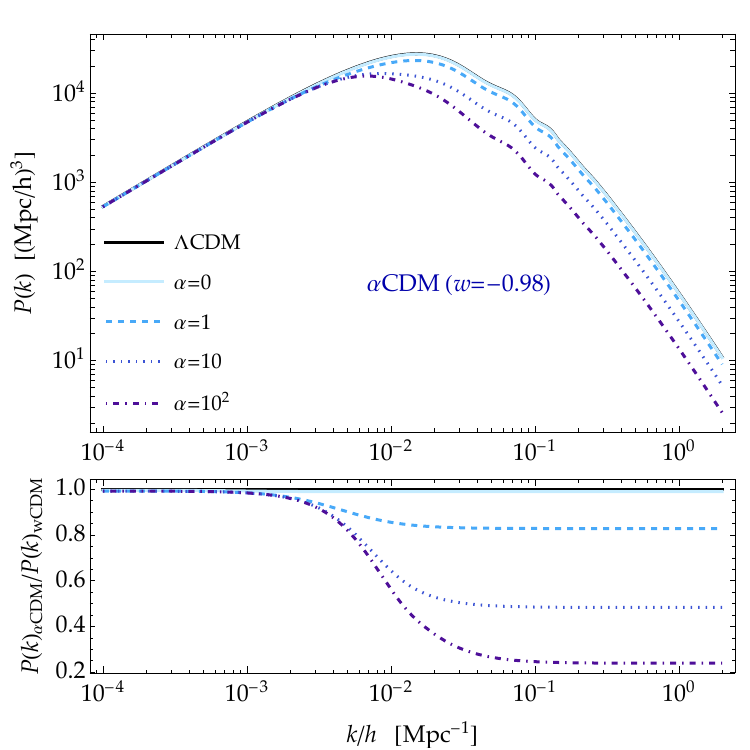}
\includegraphics[width=0.49\textwidth]{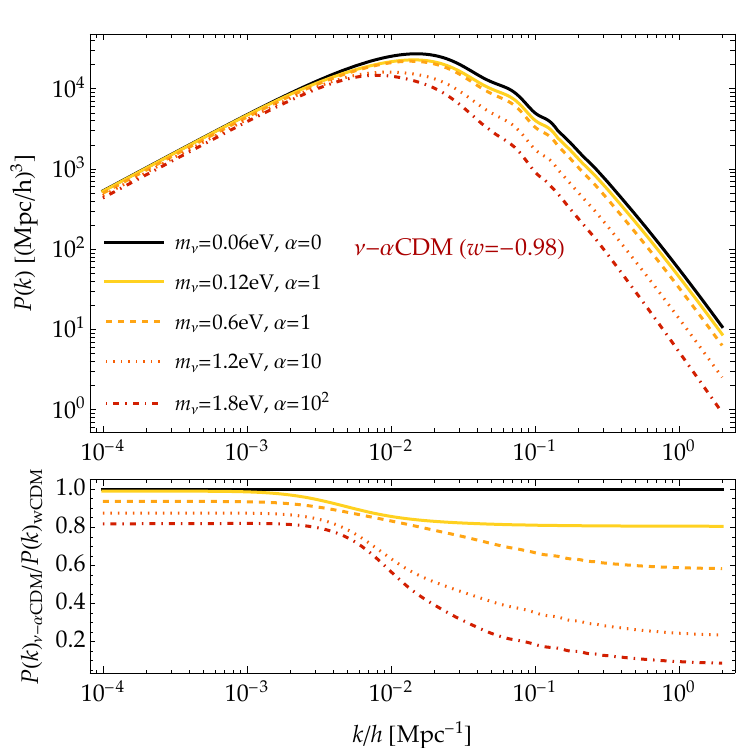}
\caption{In this plot we show the matter power spectrum for the cosmologies described in Section~\ref{sec:Results}. The top panels show the reference models $\nu - \Lambda$CDM and $\nu-w$CDM corresponding, respectively, to the $\Lambda$CDM model and the $w$CDM model ($w=-0.98$) with different values of the neutrino mass, $m_\nu$. The bottom panels illustrate the effects on the matter power spectrum for the interacting $\alpha$CDM model ($w=-0.98$) with different values for the coupling parameter $\alpha$ (left) and the joint effect of $\alpha$ and  different values of the neutrino mass, $m_\nu$. In all cases, relative variations are display w.r.t. the corresponding reference models $\Lambda$CDM and $w$CDM (with $m_\nu=0.06\,$eV and $\alpha=0$).} 
\label{Fig:Pk}
\end{figure*}

\begin{figure*}[!t]
\includegraphics[width=\textwidth]{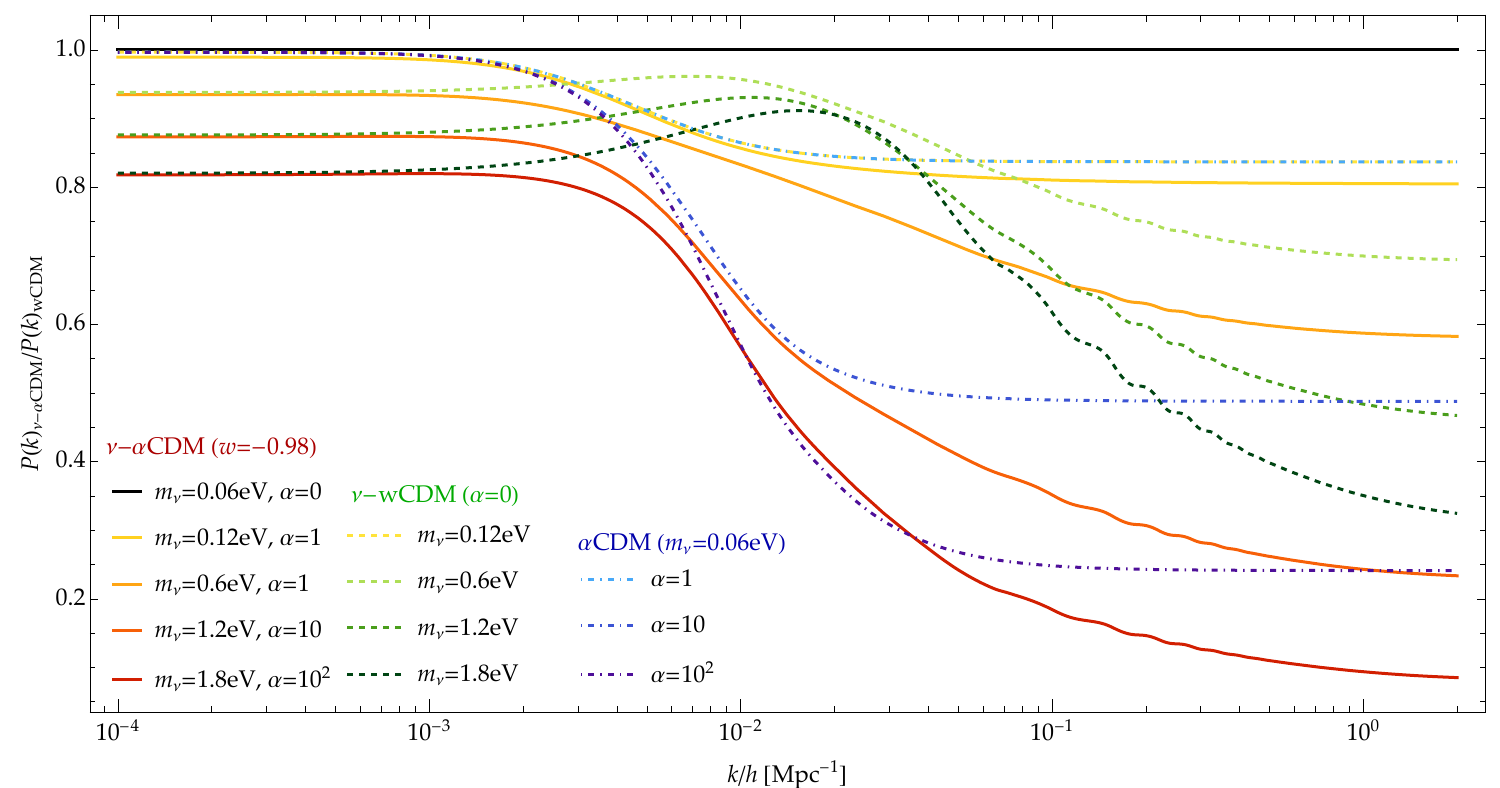}
\caption{In this plot we show the relative variations of the matter power spectrum for the studied models: dotted-dashed lines correspond to the interacting $\alpha$CDM model ($w=-0.98$ and $m_\nu=0.06\,$eV), dashed lines correspond to the $\nu-w$CDM model  (w=-0.98 and $\alpha=0$) and solid lines show the joint effect of $\alpha$ and  different values of the neutrino mass, $m_\nu$.} 
\label{Fig:PkALL}
\end{figure*}

\begin{figure*}
\includegraphics[width=0.49\textwidth]{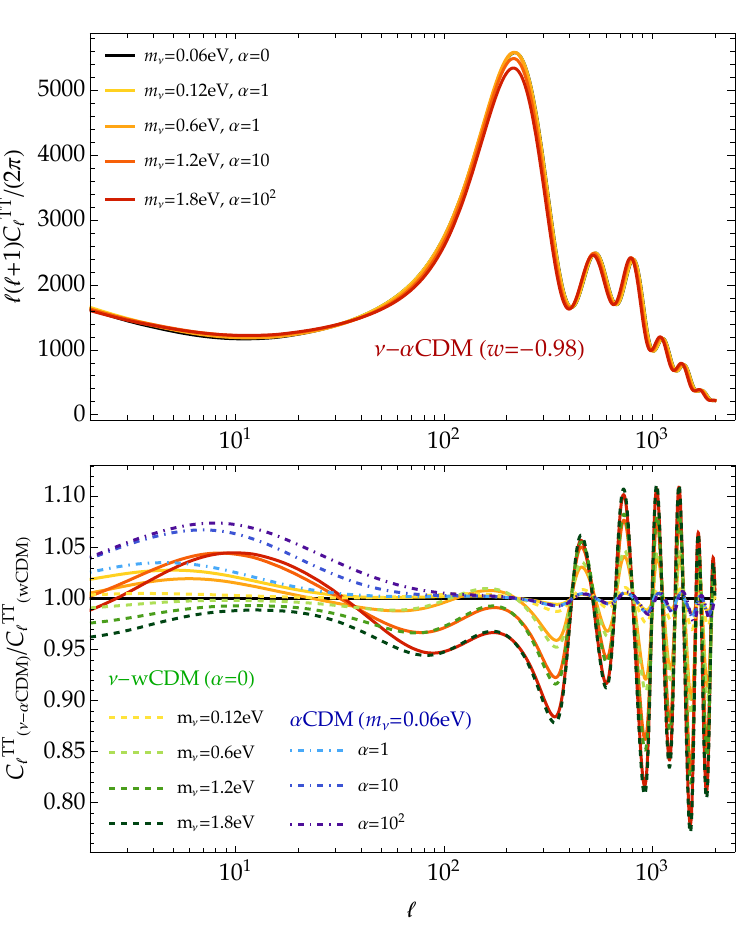}
\includegraphics[width=0.49\textwidth]{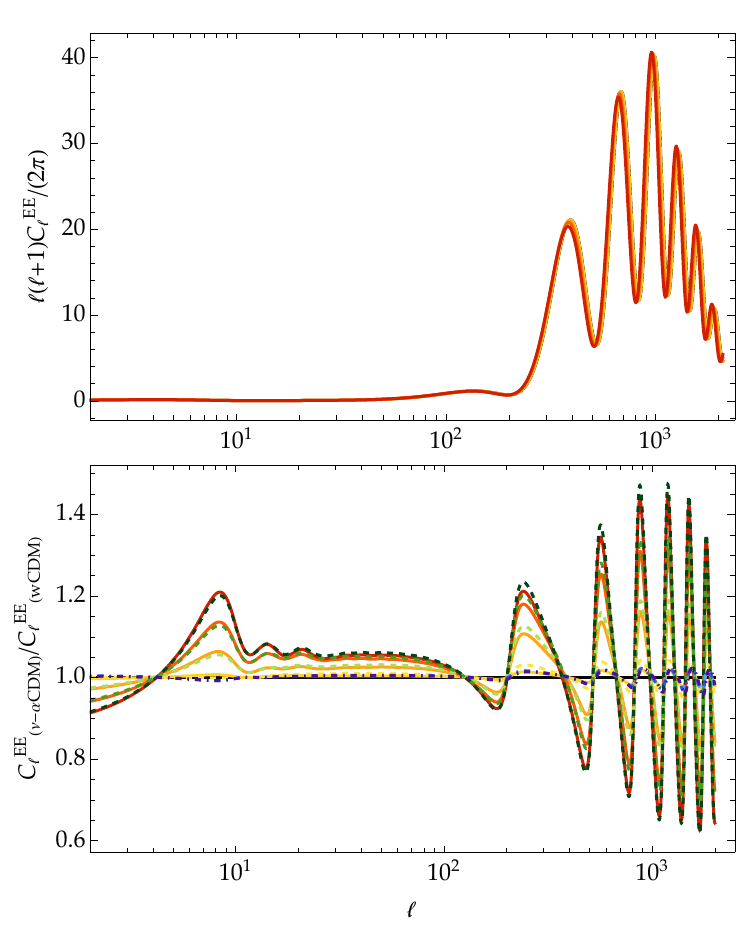} 
\includegraphics[width=0.49\textwidth]{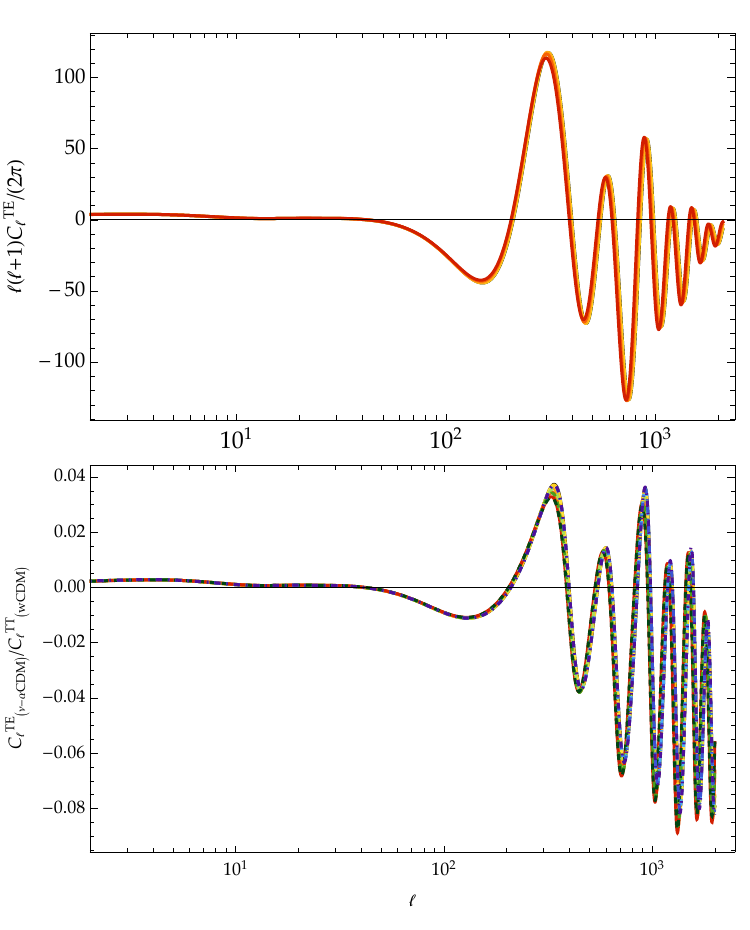}
\includegraphics[width=0.49\textwidth]{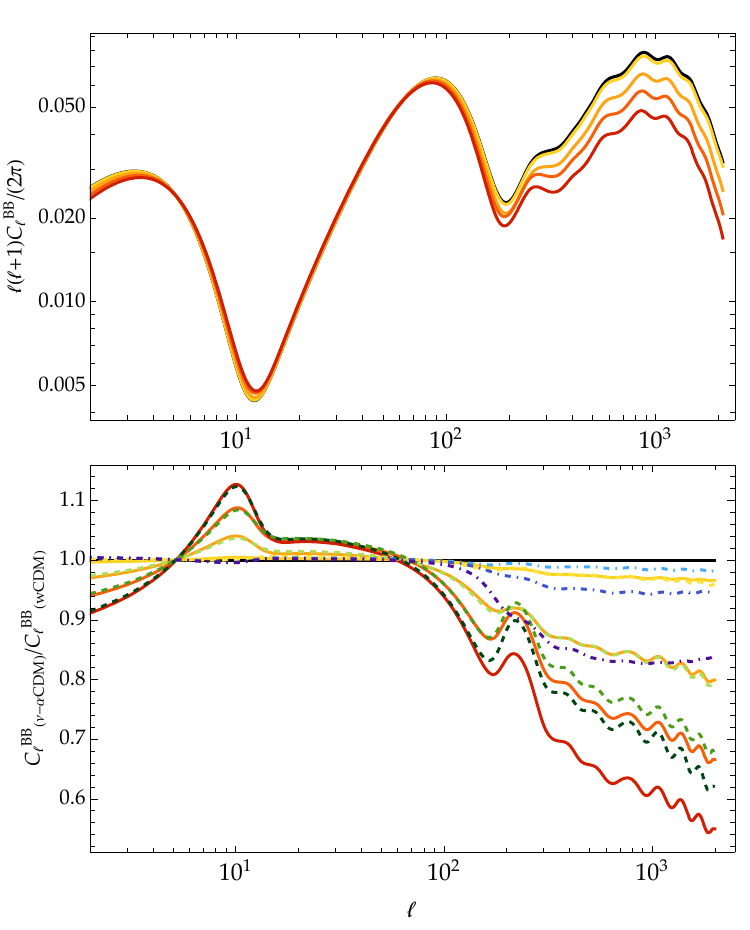}
\caption{In this Figure we display the CMB angular power spectra (temperature and polarisation) for the $\nu - \alpha$CDM model. Different combinations of the  parameters $\alpha$ and $m_\nu$ are showed for comparison. In all cases, the normalisation of relative variations is given w.r.t. the $w$CDM model ($\alpha = 0$ and $m_\nu=0.06\,$eV).} 
\label{Fig:ClsAwCDMmnu}
\end{figure*}

As advertised, the interaction simply adds a new term to the Euler equations of the coupled fluids. This term formally resembles that of the Thomson scattering between baryons and photons before decoupling. For this reason, we refer to our scenario as covariantised dark Thomson-like scattering. It is clear from the equations that the interaction requires the presence of peculiar velocities between the dark components, something that only occurs on sub-Hubble scales, as well as an interaction rate larger than the Hubble expansion rate, something that parameterically occurs when $\Gamma\gtrsim \mathcal{H}$, i.e., at late times. From this discussion, it is clear that the interaction will affect the small scales at low redshift. For those scales, the interaction provides the dark matter component with an effective pressure originated from the pressure of dark energy. This pressure will work against the dark matter clustering, thus leading to a suppression of the matter power spectrum on small scales and at late times. This suppression of the matter power spectrum looks similar to the effect of massive neutrinos when they become non-relativistic, thus it is natural to wonder if both effects could be degenerate. 

In Figure~\ref{Fig:Pk} and Figure~\ref{Fig:PkALL} we show the effects of the interaction and massive neutrinos in the matter power spectrum for different cosmologies. We can see that both the interaction and the massive neutrinos produce a similar  suppression of the matter power spectrum on small scales. However, we can already see that, while the interaction does not modify the background, the massive neutrinos cosmologies also have an impact in the background evolution and this is reflected in a modification of the power spectrum on very large scales. Of course, this can be corrected by varying other background cosmological parameters. On the other hand, the scale-dependence of the suppression of the matter power spectrum also differs in both scenarios. These different effects can help breaking the potential degeneracies between both scenarios. In particular, a full shape analysis of the power spectrum should allow to break possible degeneracies. Another distinctive feature of the elastic interaction is that it modifies the dipole of the matter power spectrum and this effect will clearly allow to distinguish the interaction from massive neutrinos, which do not produce such an effect \cite{BeltranJimenez:2022irm}. It should be clear then that the potential degeneracy between both scenarios can eventually be broken. In this work, however, we are interested in analysing if the results already obtained in the literature for the elastic interacting scenarios are prone to a degeneracy with massive neutrinos because some observables are not sensitive to the discussed effects or they are not sufficiently precise to see the differences between both scenarios. This is important because the existing studies point towards a possible detection of the interaction parameter that is significantly favoured with respect to the non-interacting scenario. It is then important to analyse if allowing for a varying neutrino mass could alter these findings.

Before proceeding to the observational constraints on the interacting scenarios with massive neutrinos, let us also comment on the effects on the CMB power spectrum. In Figure~\ref{Fig:ClsAwCDMmnu} we show the joint effects of the interaction and the neutrino mass in the cosmic microwave background. The power spectrum for temperature is most significantly modified at large scales through the late Integrated Sachs–Wolfe (ISW) effect (as expected because the interaction is relevant at very late times) and the reduction of the weak lensing effect due to the neutrino mass. The amplitude of the peaks is also modified due to the impact of the total neutrino mass in the background and perturbations evolution. 
An increase in neutrino mass also produces a decrease in the late ISW effect, reducing the CMB temperature spectrum at low $l$s. In general, we can see that the massive neutrinos have a bigger impact on the CMB than the interaction and this is in part due to the fact that the interaction does not affect the background evolution.

\section{MCMC results}
\label{sec:Results}

In order to address the possible degeneracy between the elastic interaction and massive neutrinos, we will perform Markov Chain Monte Carlo (MCMC) analyses that extend those performed in \cite{Asghari:2019qld,Figueruelo:2021elm,BeltranJimenez:2021wbq,Poulin:2022sgp} by allowing for a varying neutrino mass. For that, we will use a modified version of the Boltzmann solver for cosmological perturbations CLASS~\cite{2011arXiv1104.2932L,2011JCAP...07..034B} that includes the interaction \cite{Figueruelo:2021elm},\footnote{A modified version of CAMB~\cite{Lewis:1999bs,Howlett:2012mh} which includes the elastic interaction in the dark sector to the evolution of the cosmological perturbations is also available on request \cite{BeltranJimenez:2021wbq}. We have tested that both modified codes give consistent results.} so that we can use  the  MCMC code MontePython~\cite{Brinckmann:2018cvx,2013JCAP...02..001A} for sampling the parameter space. In our analyses we are going to consider the following two different cosmologies:
\begin{itemize}
    \item $\nu-\Lambda$CDM: the concordance model $\Lambda$CDM with two massless neutrinos and one massive neutrino with mass $m_{\nu}$, which will be a parameter in the MCMC analyses.
    \item $\nu-\alpha$CDM: the covariantised dark Thomson-like model explained in Section~\ref{sec:Covariantised} with two massless neutrinos and one massive neutrino of mass $m_{\nu}$, which again will be a parameter in the MCMC analyses together with the model parameter $\alpha$.
\end{itemize}
Consequently, the cosmological parameters to be sampled are  baryon density as  $100\Omega_{\rm b}h^2$, the dark matter density as  $\Omega_{\rm dm}h^2$, the angle of the comoving sound horizon at recombination as $100\theta_{\rm s}$, the amplitude of primordial
perturbations as $\ln(10^{10} A_{\rm s})$, the scalar spectral index $n_{\rm s}$, the optical depth $\tau_{\rm reio}$, the neutrino mass $m_\nu$ and the equation of state of dark energy $w$. In addition to that, we will also have the coupling parameter of the covariantised dark Thomson-like model $\alpha$. As derived parameters we will have the redshift of reionisation $z_{\rm reio}$, the total matter abundance $\Omega_{\rm m}$, the primordial Helium fraction $Y_{\rm He}$, the Hubble parameter $H_0$ and the root-mean-square of density fluctuations inside spheres of $8h^{-1}\;\rm{Mpc}$ radius $\sigma_8$. We will set flat priors on the parameters with bounds only for $\alpha$ and $w$ as $\alpha\in[-0.01,100]$ for the coupling parameter and $w>-1$ for the equation of state of dark energy due to stability reasons. We will also use the conservative bound on the neutrino mass of $m_\nu\in[0,2]\,\rm{eV}$.
For the datasets to be used, we will consider the following combinations:
\begin{itemize}
    \item Baseline: Planck data of the TT, TE and EE spectrum~\cite{Aghanim:2018eyx,Aghanim:2019ame}, Pantheon+ data of Supernovae Ia~\cite{Brout:2022vxf}, BAO combined  data~\cite{Beutler:2011hx,Ross:2014qpa,Hou:2020rse,Neveux:2020voa,Tamone:2020qrl,deMattia:2020fkb,duMasdesBourboux:2020pck}.
    \item Baseline+Lensing: previous data with also the Planck CMB lensing power spectrum~\cite{Aghanim:2018eyx,Aghanim:2019ame}.
    \item Baseline+Lensing+DES-Y3: previous data with also a Gaussian likelihood of the form 
    \begin{equation}
        \log\mathcal{L}_{S_{8}}=-\frac{(S_{8,\rm model}-S_{8,\rm obs})^2}{2\sigma_{S_8}^2}\;,
        \label{eq:GaussianS8}
    \end{equation}
     using the results from the DES survey third year release consisting of $S_{8,\rm obs}^{\rm DES-Y3}=0.776\pm0.017$~\cite{DES:2022urg}. The use of this data has some caveats (see e.g. the discussion in this respect in Ref.~\cite{BeltranJimenez:2021wbq}), but it is a common approach and we will also adopt it here. This is however a very important point because it is precisely this data and used in this manner what permits to constrain the interaction parameter. We could have also introduced Sunyaev-Zeldovich data~\cite{Planck:2015lwi} or KIDS data~\cite{Heymans:2020gsg} in the same way, but it would not change substantially our results and it will not be necessary for our purpose here, which is to show the impact of varying the neutrino mass.
\end{itemize}
In all the scenarios, we will make use of the Gelman-Rubin criteria~\cite{10.1214/ss/1177011136} satisfying that $\vert R-1\vert < 0.01$ in order to ensure the convergence of the chains.\\

\begin{figure*}[!t]
\includegraphics[width=0.9\textwidth]{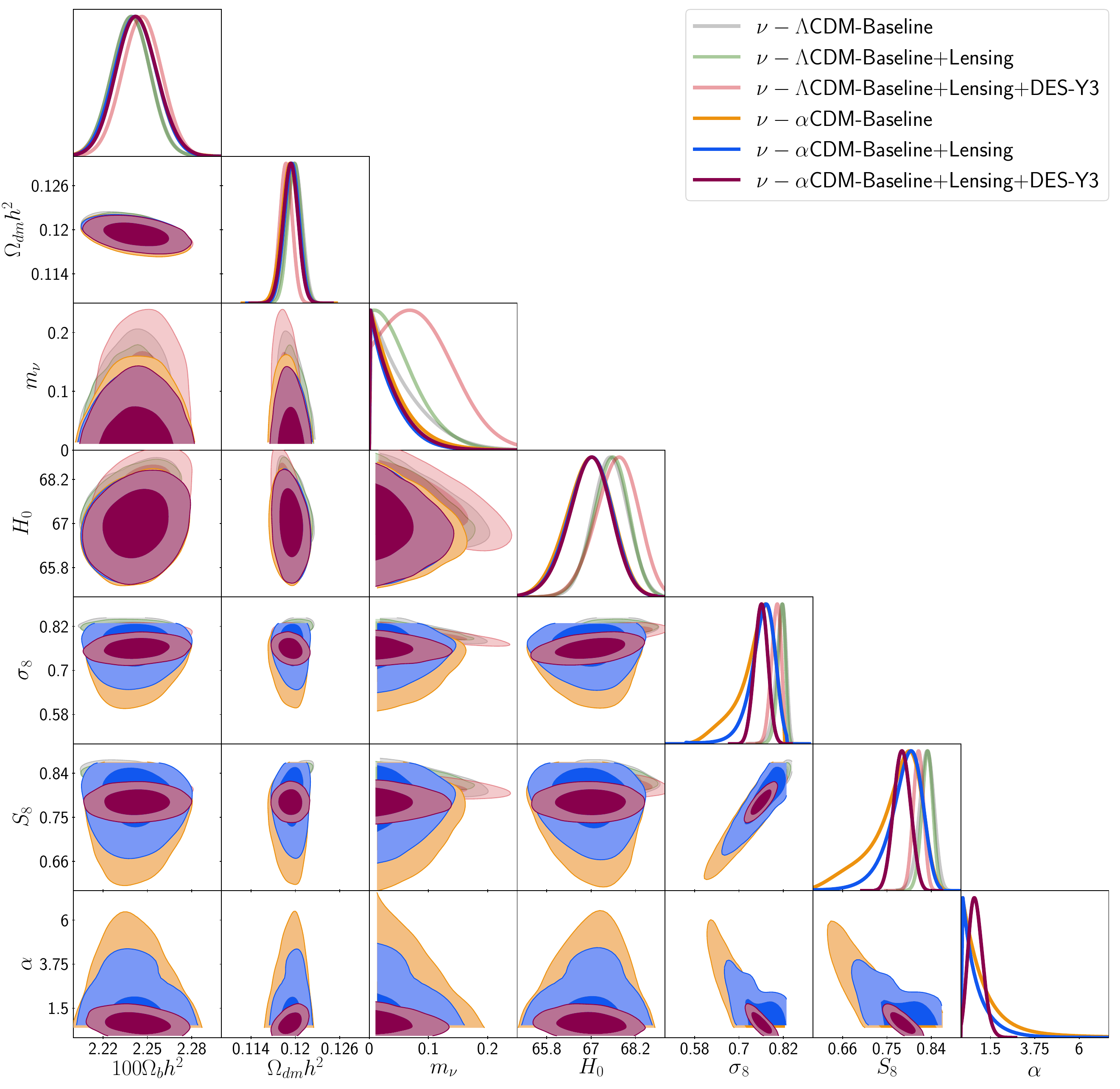}
\caption{The one-dimensional posterior distributions and the two-dimensional contours obtained for some relevant parameters using all the different data sets considered for the studied models: $\nu-\Lambda$CDM which corresponds to  the concordance model $\Lambda$CDM with one massive neutrino of mass $m_\nu$ and $\nu-\alpha$CDM which corresponds to  the interacting model $\alpha$CDM with one massive neutrino of mass $m_\nu$.} 
\label{Fig:MCMC_triangle}
\end{figure*}

The results that we obtain are given in Table~\ref{tab:MCMC_results} and in Figure~\ref{Fig:MCMC_triangle} where we show the constraints for some relevant parameters in the $\nu-\Lambda$CDM and $\nu-\alpha$CDM scenarios. A general conclusion inferred from the results is that, as expected by the very nature of the interaction, most of the parameters remain unchanged in the presence of the momentum transfer. The other outcome from the MCMC analyses is that the interaction can only be detected once we add low redshift data, in our case the third year DES data, in agreement with previous findings \cite{Figueruelo:2021elm,BeltranJimenez:2021wbq,Poulin:2022sgp}. As a matter of fact, when no low-redshift data is used, the results show no lower constraint for the coupling parameter and we only obtain the bounds:
\begin{eqnarray}  
    \alpha&<{1.23}^{+0.11} {}^{+3.10}\;\; &\text{(CMB-Planck+Pantheon}\nonumber\\
    & &\text{+BAO)}\;,     \\
    \alpha&<{0.92}^{+0.21} {}^{+2.0}\;\; &\text{(CMB-Planck+Pantheon}\nonumber\\
    & &\text{+Lensing+BAO)}\;.
\end{eqnarray}
However, when we add the low-redshift data in the very simple form of the Gaussian likelihood explained above, we are able to establish both a lower and an upper constraint on the coupling parameter of the covariantised dark Thomson-like scattering, resulting in the following value:
\begin{eqnarray}  
    \alpha=&{0.76}_{-0.41}^{+0.31} {}_{-0.68}^{+0.71}\;\; &\text{(CMB-Planck+Pantheon}\nonumber\\
    & &\text{+Lensing+BAO}\nonumber\\
    & &\text{+DES-Y3)}\;.
\end{eqnarray}
Let us notice that the coupling parameter is detected to be different from zero with a $2\sigma$ confidence level only once the low-redshift data are used.  Similar occurrences have been found using other datasets, such as the CFHTLenS~\cite{CFHTL} or the Planck Sunyaev–Zeldovich clusters counts~\cite{Planck:2015lwi}. In order to understand the previous result, one has to first consider that the covariantised dark Thomson-like scattering is a very late-Universe interaction and only occurs on small scales, while the previous surveys are not, except for DES-Y3. Therefore, they are unable to set strong constraints on the strength of the interaction as they are not sensitive to it, i.e., both high redshift perturbations and the background cosmology are oblivious to the interaction. The second reason for this trend relates to the nature of the interaction and the $\sigma_8$/$S_8$ tension. Those previous surveys, used in the very simple form of a Gaussian likelihood, capture the low-redshift indication for less structures in our current Universe than the amount of structures suggested by early Universe datasets. Therefore, a low-redshift mechanism capable of reducing the clustering, as the momentum transfer, will naturally accommodate those values for $S_8$, as we have in turn obtained. 

\begin{figure*}
\includegraphics[width=0.4\textwidth]{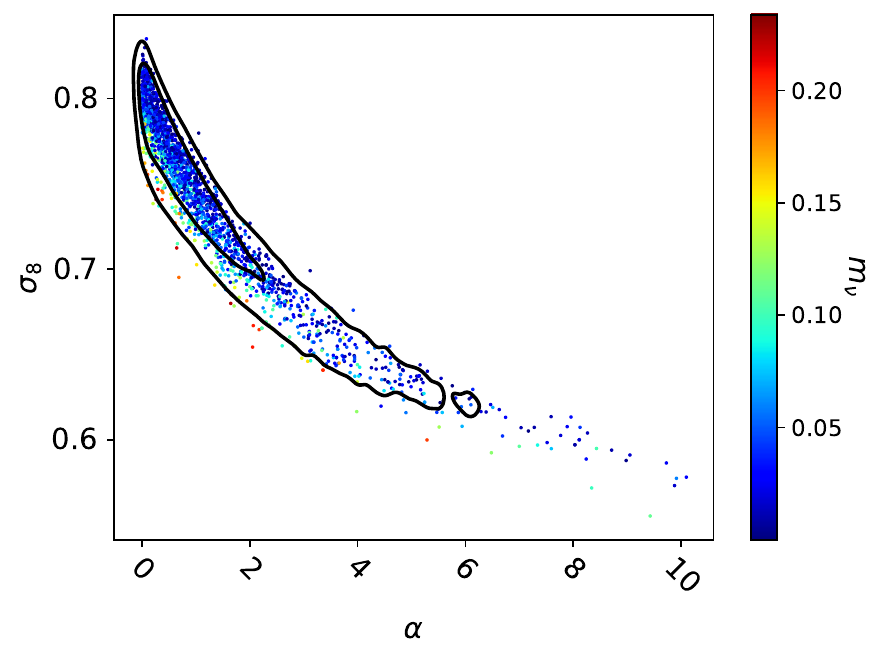}
\includegraphics[width=0.4\textwidth]{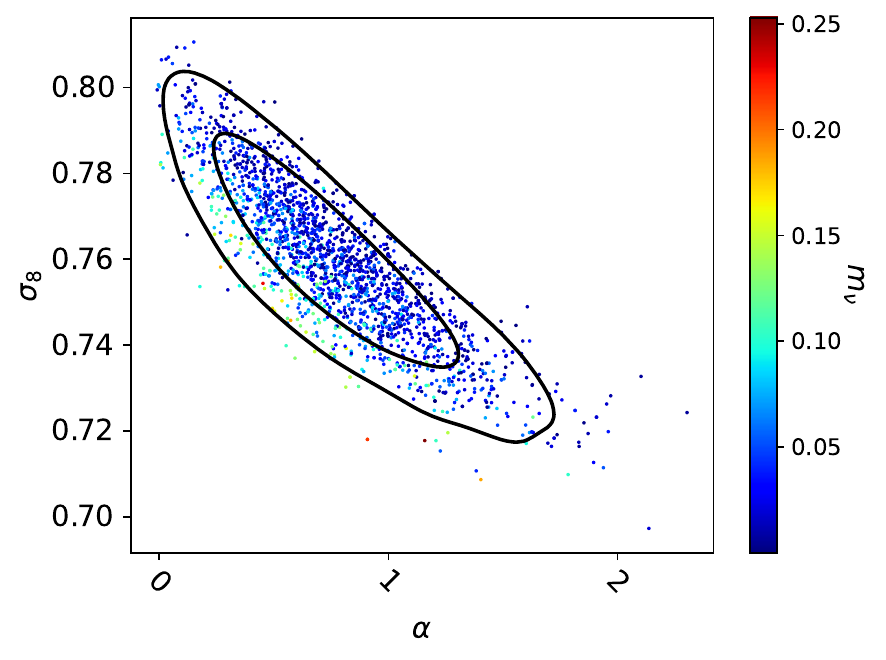}
\caption{$\nu-\alpha$CDM plane $\alpha-\sigma_8$ with $m_\nu$ colour map: Baseline (left) and Baseline+Lensing+DES-Y3 (right). From this figure we see how the neutrino mass does not play any role in the value of $\sigma_8$ which is only correlated with the interaction parameter.} \label{Fig:alpha:alpha-sigma8vsmnu}
\end{figure*}

\begin{figure*}
\includegraphics[width=0.4\textwidth]{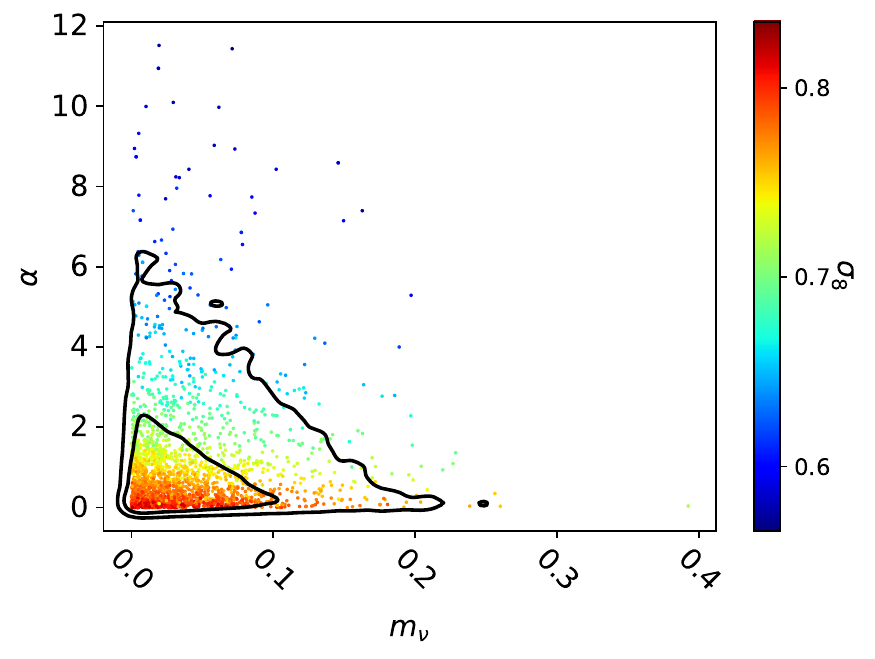}
\includegraphics[width=0.4\textwidth]{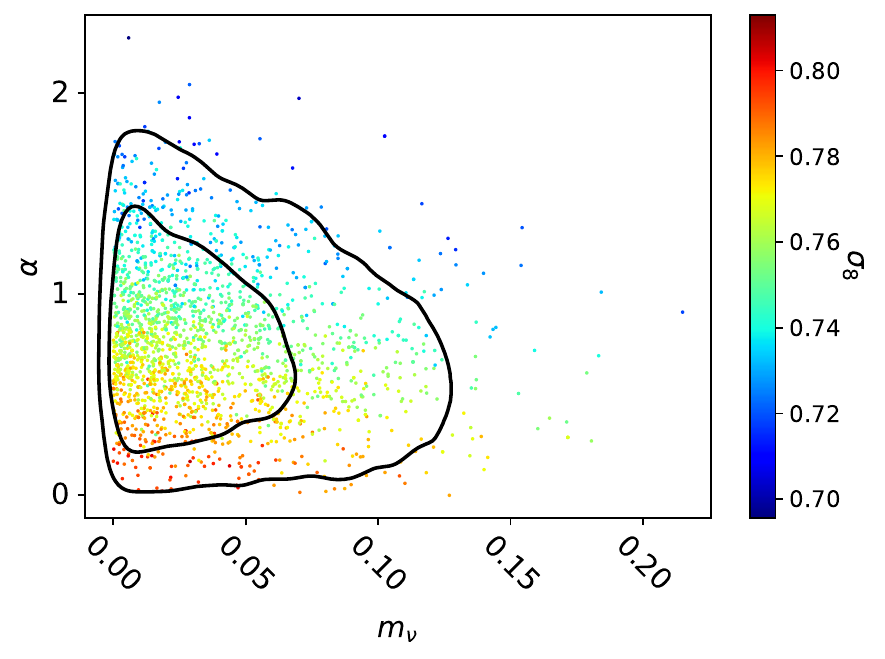}
\caption{$\nu-\alpha$CDM plane $\alpha-\sigma_8$ with $m_\nu$ colour map: Baseline (left) and Baseline+Lensing+DES-Y3 (right). This figure further illustrates how the neutrino mass does not play any role in having lower values of $\sigma_8$. We also see how the addition of DES-Y3 permits to constrain $\alpha$.} \label{Fig:alpha:alpha-mnuvssigma8}
\end{figure*}

These results permit us to answer our question, namely: there is no strong degeneracy between the interaction and the mass of the neutrinos so that allowing for a varying neutrino mass does not degrade the potential detection of the interaction. This is clearly seen in the $\alpha-m_\nu$ plane in Figure~\ref{Fig:MCMC_triangle} where we see that the constraint on $\alpha$ is insensitive to the value of $m_\nu$. Remarkably, we see that, although more massive neutrinos could lower the value of $\sigma_8$ (or $S_8$), it is the effect of the interaction what mainly drives its value. In fact, we can see the clear correlation between $\alpha$ and $\sigma_8$, while the values of $\sigma_8$ and $m_\nu$ are not correlated. In order to illustrate this more clearly, in Figure \ref{Fig:alpha:alpha-sigma8vsmnu} we show the $\sigma_8-\alpha$ plane coloured with the value of $m_\nu$. In that figure it is apparent that the mass of the neutrinos plays no role in the value of $\sigma_8$ and that it is exclusively the interaction what drives $\sigma_8$ towards smaller values. This is also shown in Figure~\ref{Fig:alpha:alpha-mnuvssigma8} where the plane $\alpha-m_\nu$ coloured with the value of $\sigma_8$ is shown.

\begin{figure*}
\includegraphics[width=0.4\textwidth]{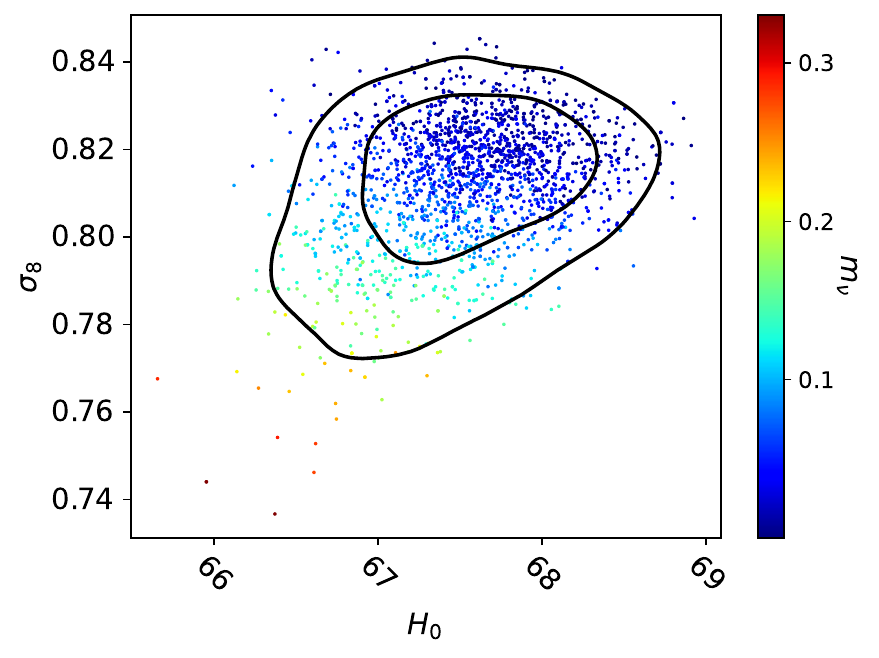}
\includegraphics[width=0.4\textwidth]{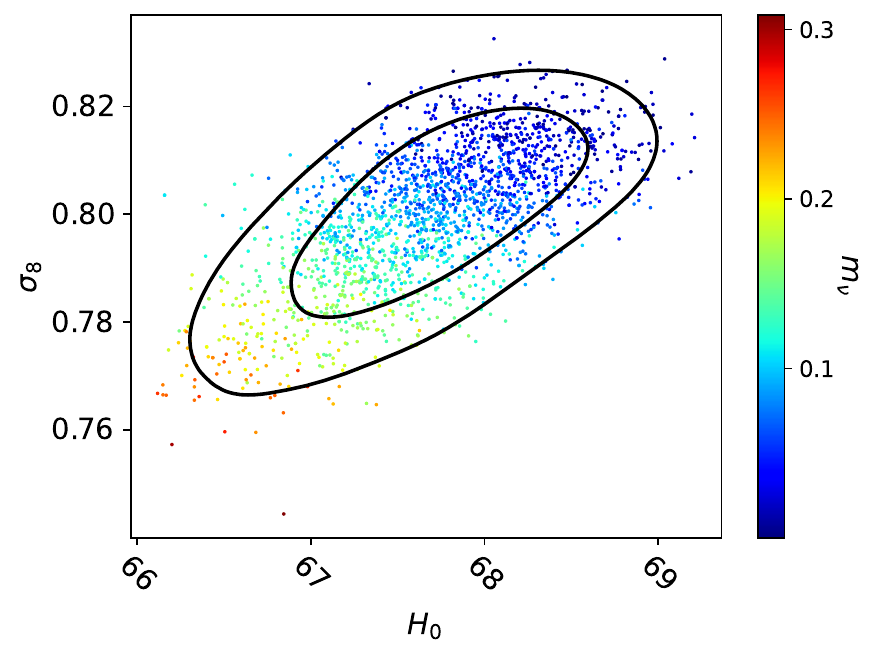}
\caption{$\nu-\Lambda$CDM plane $\sigma_8-H_0$ with $m_\nu$ colour map: Baseline and Baseline+Lensing+DES-Y3. We can see how lower values of $\sigma_8$ come at the expense of also lowering $H_0$ so improving one of the tensions worsen the other. Furthermore, the colour map shows the correlation between the neutrino mass and the value of $\sigma_8$.} 
\label{Fig:LCDM:s8-H0vsmnu}
\end{figure*}

\begin{figure*}
\includegraphics[width=0.4\textwidth]{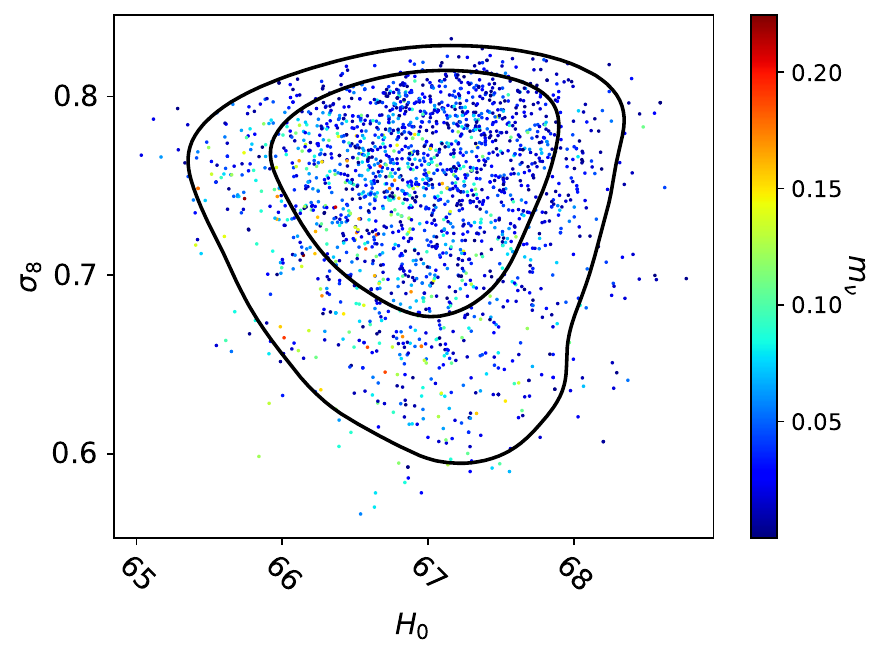}
\includegraphics[width=0.4\textwidth]{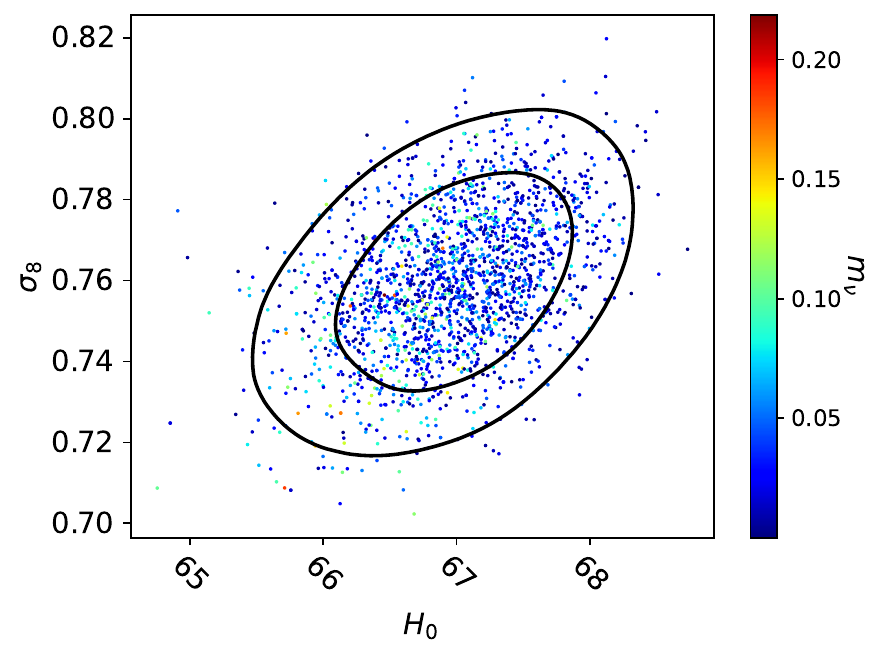}
\caption{$\nu-\alpha$CDM plane $\sigma_8-H_0$ with $m_\nu$ colour map: Baseline (left) and Baseline+Lensing+DES-Y3 (right). In this Figure we see how the correlation between $\sigma_8$ and $H_0$ disappears as well as the correlation between $\sigma_8$ and $m_\nu$.} 
\label{Fig:alpha:s8-H0vsmnu}
\end{figure*}

\begin{figure*}
\includegraphics[width=0.4\textwidth]{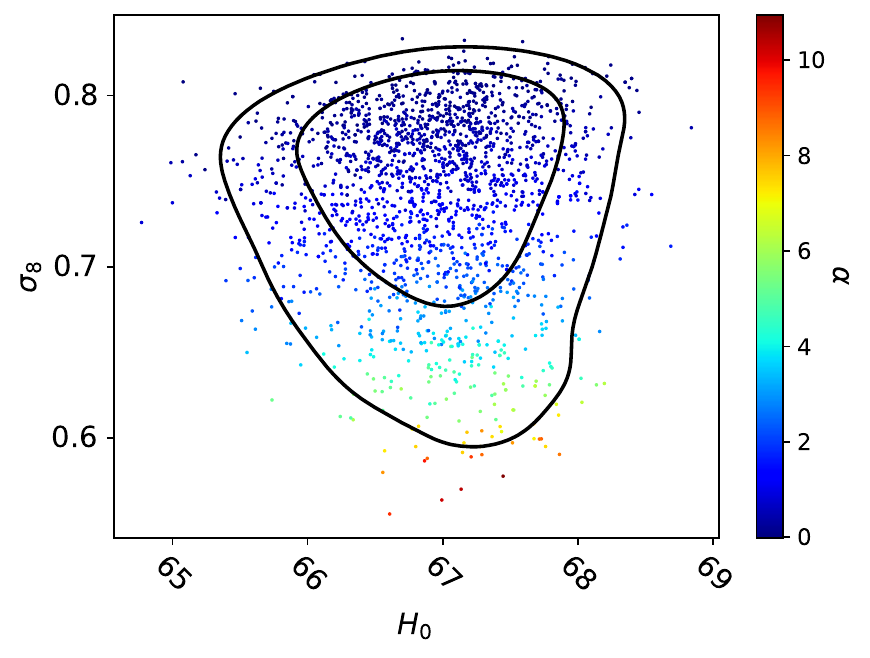}
\includegraphics[width=0.4\textwidth]{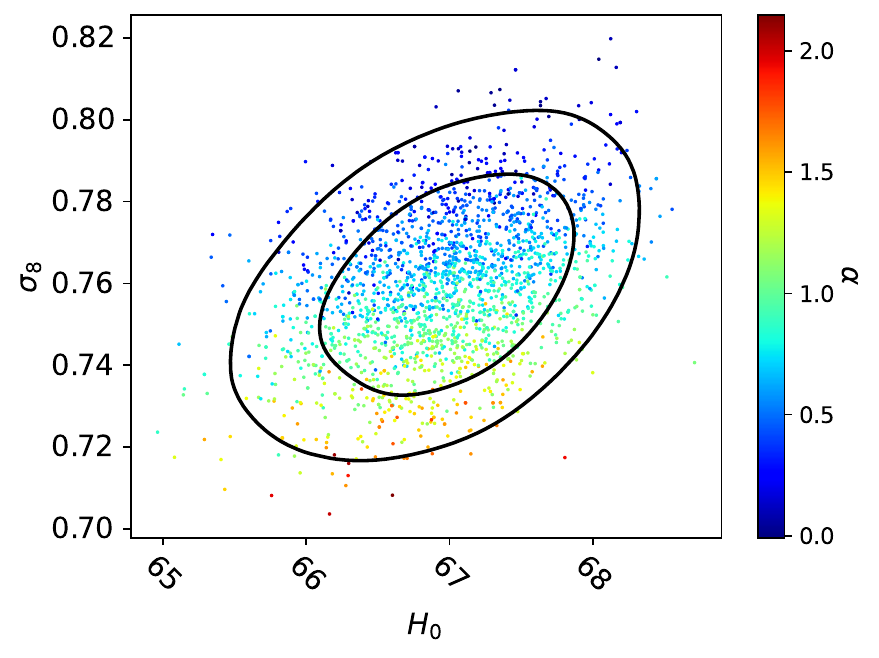}
\caption{$\nu-\alpha$CDM plane $\sigma_8-H_0$ with $\alpha$ colour map: Baseline (left) and Baseline+Lensing+DES-Y3 (right). In this figure we can confirm that the lower values of $\sigma_8$ are driven by having higher values of $\alpha$.} \label{Fig:alpha:s8-H0vsalpha}
\end{figure*}

Apart from this, there are certain parameters that deserve further explanations:
\begin{itemize}
    \item $H_0$: the value of the Hubble constant is found to be consistently lower for the $\nu-\alpha$CDM model compared to $\nu-\Lambda$CDM one, although it is not strongly significant as it is within the $1\sigma$ level. The reason for that cannot  be linked to the covariantised dark Thomson-like scattering per se since the interaction does not change, by its pure nature, the background cosmology  to where $H_0$ belongs. The reason is rather the fact that the interacting scenario requires a value of the dark energy equation of state slightly bigger than $-1$, while the $\nu-\Lambda$CDM has $w=-1$ and this is what ultimately leads to the slightly lower value of the Hubble constant. 
    
    \item $m_\nu$: although smaller upper constraints are found when the interaction is taken into account, they are not very significant. A remarkable feature of the interaction is that the typical correlation in the $\sigma_8$-$H_0$ plane with the mass of the neutrino $m_\nu$ disappears when we introduce the interaction. This can be clearly seen when comparing Figure~\ref{Fig:LCDM:s8-H0vsmnu} and Figure~\ref{Fig:alpha:s8-H0vsmnu}. The scenario without interaction exhibits a correlation so that higher neutrino masses lead to smaller values of $\sigma_8$ (as expected), but this comes at the expense of also lowering the value of $H_0$. However,  when the interaction is turned on we see two distinctive effects. Firstly, the correlation between $\sigma_8$ and $H_0$ disappears and secondly the mass of the neutrinos ceases being correlated with lower values of $\sigma_8$. This represents a very remarkable feature of the elastic interaction since it completely disentangle the mass of the neutrinos from the value of $\sigma_8$.

    \item $\sigma_8/S_8$: the interaction consistently gives rise to a lower value for both $\sigma_8$ and $S_8$ parameters as a result of the momentum transfer. Moreover, it is important to notice that the interaction allows for an alleviation of the $\sigma_8$ tension without worsening the $H_0$ tension as inferred from the $\sigma_8-H_0$ plane. The correlation between $\sigma_8$ and the coupling parameter is clearly reflected in Figure~\ref{Fig:alpha:s8-H0vsalpha} and Figure~\ref{Fig:alpha:alpha-sigma8vsmnu}, where we can see that the larger the value of $\alpha$, the smaller the value of $\sigma_8$. This is clearly explained by the nature of the interaction, since it prevents the clustering. Again in Figure~\ref{Fig:alpha:alpha-mnuvssigma8}, we clearly see that the value of $\sigma_8$ is now almost insensitive to the neutrino mass $m_\nu$ once we introduce the interaction.
\end{itemize}

\begin{figure*}[!t]
\includegraphics[width=0.9\textwidth]{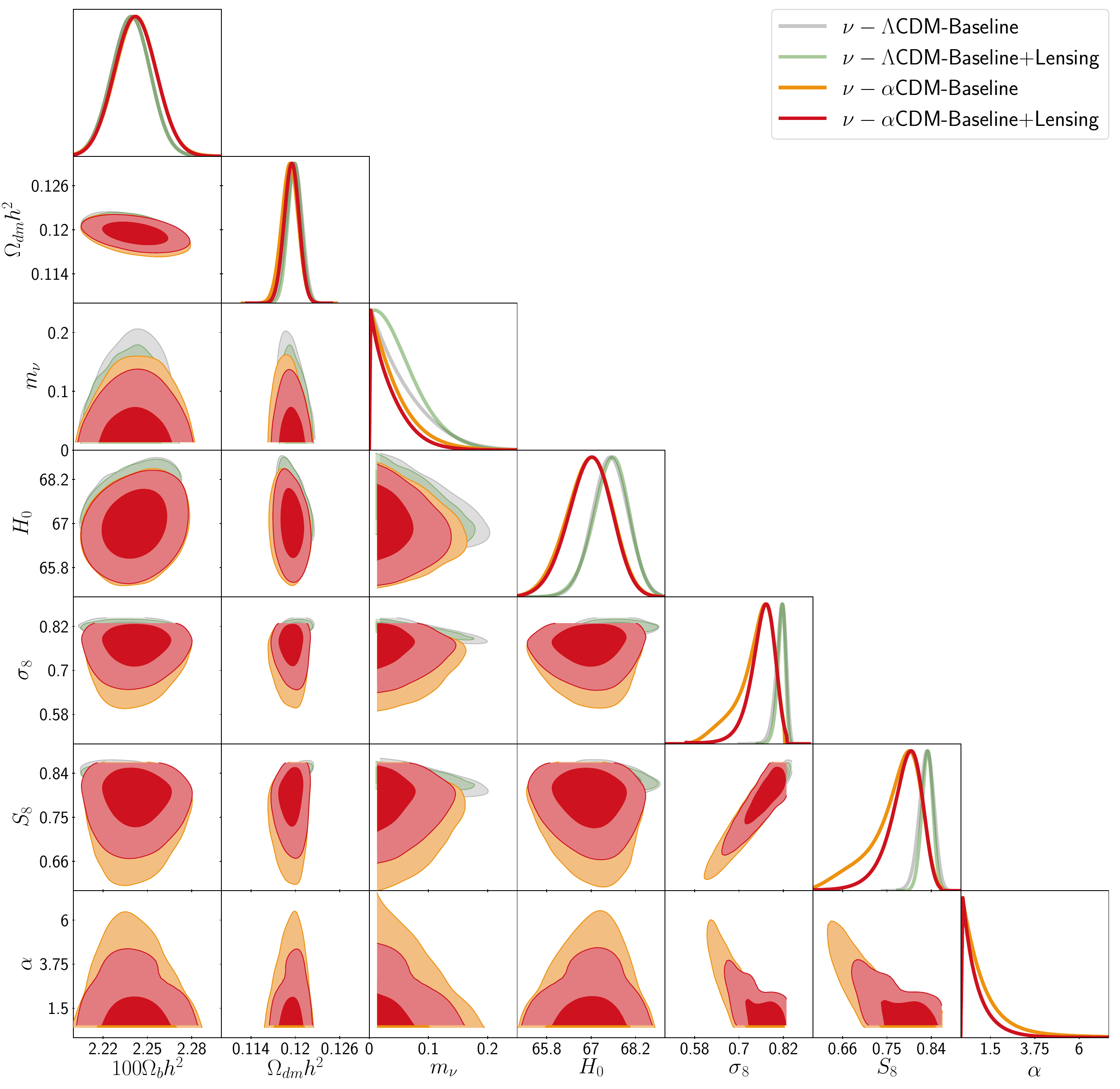}
\caption{The one-dimensional posterior distributions and the two-dimensional contours obtained for some relevant parameters using different data sets, except the DES-Y3 one, for the studied models: $\nu-\Lambda$CDM which corresponds to  the concordance model $\Lambda$CDM with one massive neutrino of mass $m_\nu$ and $\nu-\alpha$CDM which corresponds to  the interacting model $\alpha$CDM with one massive neutrino of mass $m_\nu$.} 
\label{Fig:fit_no_low_z}
\end{figure*}

In addition to the parameter constraints that we have discussed, it remains to analyse the goodness of the fit. In Table~\ref{tab:MCMC_results} we provide the best-fit values of $\chi^2$ together with the AIC~\cite{Akaike:AIC} information criteria. Comparing the results to the cases with only massless neutrinos or with one massive neutrino fixed to $m_\nu=0.06\,$eV displayed on Table~\ref{tab:MCMC_results_vis}, the preferences are weakened, as expected because we have fewer free parameters and, then, the AIC criteria penalises less. However,  the important result to highlight is that $\nu-\alpha$CDM is preferred over $\nu-\Lambda$CDM when we add the low-redshift (DES-Y3) data. Thus, this means that even when we allow for the neutrino masses to vary, the interacting scenario is favoured over the non-interacting scenario.

\begin{table*}
\begin{center}
\renewcommand{\arraystretch}{1.8}
\begin{tabular}{||c|c|c|c||}
	\hline
	\hline
	   {\bf $\nu-\Lambda$CDM }    & Baseline  & Baseline+Lensing & Baseline+Lensing+DES-Y3  \\
	\hline 
	\hline 
 $H_0$ [km/s/Mpc]  & ${67.529}_{-0.45}^{+0.55}{}_{-1.01}^{+0.97}$ &  ${67.52}_{-0.48}^{+0.53} {}_{-1.0}^{+1.0}$ & ${67.69}_{-0.55}^{+0.64}{}_{-1.19}^{+1.13}$ 
 \\ \hline 
 $m_{\nu}$ & $ <{0.055}^{+0.013} {}^{+0.100}$    &$ <{0.053}^{+0.014} {}^{+0.084}$  & $<{0.088}^{+0.031} {}^{+0.107}$ \\ \hline
 $\sigma_8$  & ${0.813}_{-0.010}^{+0.017}{}_{-0.030}^{+0.027}$ &  ${0.814}_{-0.009}^{+0.013} {}_{-0.024}^{+0.021}$ & ${0.800}_{-0.011}^{+0.016}{}_{-0.0274}^{+0.024}$
 \\ \hline 
 $S_8$ & ${0.830}_{-0.014}^{+0.017}{}_{-0.030}^{+0.0274}$ &  ${0.832}_{-0.011}^{+0.012} {}_{-0.024}^{+0.023}$ & ${0.814}_{-0.011}^{+0.011}{}_{-0.022}^{+0.021}$
 \\ \hline  
 $\chi^2_{\rm best\;fit}$ &  4196  &   4205 &4212  \\ \hline
\hline
	   {\bf$\nu-\alpha$CDM }    & Baseline  & Baseline+Lensing & Baseline+Lensing+DES-Y3 \\ 
	\hline 
	\hline 
 $H_0$ [km/s/Mpc] & ${66.943}_{-0.60}^{+0.68}{}_{-1.29}^{+1.23}$ &  ${66.96}_{-0.59}^{+0.65} {}_{-1.27}^{+1.19}$ & ${66.96}_{-0.58}^{+0.66}{}_{-1.24}^{+1.23}$
 \\ \hline 
 $m_{\nu}$ & $<{0.041}^{+0.009} {}^{+0.081}$    &$ <{0.035}^{+0.008} {}^{+0.068}$ &   $<{0.037}^{+0.008} {}^{+0.069}$ \\ \hline
 $\sigma_8$ & ${0.744}_{-0.028}^{+0.064 }{}_{-0.109}^{+0.079}$ &  ${0.762}_{-0.021}^{+0.051} {}_{-0.086}^{+0.065}$ &  ${0.760}_{-0.019}^{+0.019}{}_{-0.036}^{+0.036}$
 \\ \hline 
 $S_8$ & ${0.765}_{-0.032}^{+0.065}{}_{-0.113}^{+0.089}$ &  ${0.784}_{-0.024}^{+0.052} {}_{-0.088}^{+0.073}$& ${0.781}_{-0.017}^{+0.017}{}_{-0.034}^{+0.34}$
 \\ \hline 
 $\alpha$  & $<{1.23}^{+0.11} {}^{+3.10}$    &$ <{0.92}^{+0.21} {}^{+2.0}$   & ${0.76}_{-0.41}^{+0.31} {}_{-0.68}^{+0.71}$\\ \hline
$\chi^2_{\rm best\;fit}$ & 4195  &  4203 & 4205\\ \hline
 $\Delta$AIC & 3.3 & 1.2  & -2.7 \\ \hline 
\hline 
\end{tabular}
\end{center}
\caption{Mean likelihood values and $1\sigma$ and $2\sigma$ limits for some relevant parameters using different data sets for the studied models: $\nu-\Lambda$CDM which corresponds to  the concordance model $\Lambda$CDM with one massive neutrino of mass $m_\nu$ and $\nu-\alpha$CDM which corresponds to  the interacting model $\alpha$CDM with one massive neutrino of mass $m_\nu$.} 
\label{tab:MCMC_results}
\end{table*}

\begin{table*}
\begin{center}
\renewcommand{\arraystretch}{1.8}
\begin{tabular}{||c|c|c||}
	\hline
	\hline
	   {\bf  }     & one massive $\nu$ & massless $\nu$ \\
	\hline 
	\hline 
 $H_0$ [km/s/Mpc]   & ${68.00}_{-0.51}^{+0.61}{}_{-1.15}^{+1.09}$ & ${67.09}_{-0.57}^{+0.67}{}_{-1.23}^{+1.19}$ 
 \\ \hline 
 $\sigma_8$   & ${0.758}_{-0.017}^{+0.019}{}_{-0.035}^{+0.035}$ & ${0.762}_{-0.017}^{+0.018}{}_{-0.036}^{+0.035}$
 \\ \hline 
 $S_8$  & ${0.781}_{-0.016}^{+0.017}{}_{-0.034}^{+0.033}$ & ${0.781}_{-0.017}^{+0.017}{}_{-0.033}^{+0.034}$
 \\ \hline  
 $\alpha$    & ${0.72}_{-0.41}^{+0.29} {}_{-0.67}^{+0.70}$ & ${0.81}_{-0.41}^{+0.30} {}_{-0.69}^{+0.72}$
 \\ \hline
 $\chi^2_{\rm best\;fit}$ &  4207 & 4205\\ \hline
 $\Delta$AIC & -3.1 & -4.9 \\ \hline 
 \hline
\end{tabular}
\end{center}
\caption{Mean likelihood values and $1\sigma$ and $2\sigma$ limits for some relevant parameters using for $\nu-\alpha$CDM but fixing the neutrino mass to standard value $m_\nu=0.06\;\text{eV}$ and simple $\alpha$CDM with massless neutrinos.} 
\label{tab:MCMC_results_vis}
\end{table*}

\begin{table*}
\begin{center}
\renewcommand{\arraystretch}{1.8}
\begin{tabular}{||c|c|c|c|c||}
	\hline
	\hline
	    Case  & $\alpha$ mean & $<1\sigma$ bound &  $<2\sigma$ bound& $<3\sigma$ bound\\
	\hline 
	\hline 
    SZ data, $m_v=0.06$  & $1.005$ & $0.673$ &  $0.437$ & $0.249$\\
	\hline
 DES data, $m_v=0.00$  & $0.81$ & $0.40$ &  $0.12$ & $0.00197$\\
	\hline 
 DES data, $m_v=0.06$  & $0.72$ & $0.31$ &  $0.05$ & $-0.00989$\\
	\hline 
 DES data, $m_v$ free, constraint to $m_v<0.037^{+0.008}$  & $0.76$ & $0.35$ &  $0.08$ & $-0.0092$\\
	\hline 
 Ref.~\cite{Poulin:2022sgp} ($m_v=0.06$ and DES+other data)  & $0.82$ & $0.46$ &  - & -\\
	\hline 
 \hline
\end{tabular}
\end{center}
\caption{Summary results.} 
\label{tab:Summary_results}
\end{table*}

\section{Conclusions}
\label{sec:Conclusions}

In this paper we have investigated the possible presence of a degeneracy between the neutrino mass and the coupling parameter of a momentum transfer interaction in the realm of the covariantised dark Thomson-like scattering. This model has been investigated in previous works and it has been shown to be a promising scenario for alleviating the $\sigma_8$ tension and, furthermore, the addition of low-redshift data seems to signal the presence of the interaction. Since these scenarios give a suppression of the matter power spectrum on small scales and this effect is to some extent shared by scenarios with massive neutrinos, it is important to unveil whether there are degeneracies between both effects.

We have first studied the effects of the interaction and the neutrino mass in standard observables like the matter power spectrum and the cosmic microwave background. We have shown how the small scale suppression has a different scale dependence in both cases, being therefore a first probe of the non-existence of the $\alpha-m_\nu$ degeneracy. Furthermore, as we have argued, the massive neutrinos also affect the background cosmology, while the elastic interaction, by construction, leaves the background unaffected and this could be another way of breaking the degeneracies. However, there are observables that are not very sensitive to the different scale dependence of the suppression and the modifications on the background evolution could be compensated with variations of other background quantities and, thus, there could still be some residual degeneracies.

In order to set clearly the existence or not of the degeneracy we have performed several MCMC analyses using the latest available datasets. In particular, cosmic microwave background and baryonic acoustic oscillations data, while we also use the results from DES third year for the $S_8$ parameter as a Gaussian prior in our analyses. As already found in previous studies, the interaction can only be detected when the latest dataset, DES-Y3, is considered. Without that dataset, or other low-redshift probes, we can only put an upper bound on the coupling parameter of the covariantised dark Thomson-like scattering. Once low-redshift information is added, we find a {\it detection} of the interaction at more than $2\sigma$ confidence level. Such detection is intrinsically related to a lower value of the $\sigma_8$ (or $S_8$) parameter, since the interaction induces a suppression of structures which is precisely captured by that parameter. Low-redshift datasets are continuously suggesting there are less structures in our current Universe than the expected ones from early Universe probes like the cosmic microwave background. Consequently, a late-time interaction, like the one studied here, can naturally accommodate both early and late Universe probes to a compatible value for the $\sigma_8$ or $S_8$ parameters and, thus, solving the corresponding $\sigma_8$ or $S_8$ tension. The other cosmological parameters remain oblivious to the presence of the interaction.

Regarding the possible degeneracy between the neutrino mass and the coupling parameter, we have been able to establish in a clear manner that such a degeneracy does not exist. Instead, we discovered that the well-known correlation in the $\sigma_8-H_0$ plane with the neutrino mass disappears once the interaction is considered. Given this interaction is able to alleviate or even solve the $\sigma_8$ or $S_8$ tension, one can envision a mechanism to solve the $H_0$ tension which combined with the momentum transfer models could simultaneously account for both tensions. This particular situation is something that does not happen in the standard $\Lambda$CDM model or in a broad plethora of alternative cosmological descriptions. As a final comment, let us emphasise that our elastic interacting scenario serves as a proxy for more general cosmologies featuring a pure momentum exchange, so our findings are expected to be also valid for those scenarios. 

{\bf Acknowledgements:} We thank Dario Bettoni for useful discussions and collaboration in related works. JBJ, DF and FATP acknowledge support from the {\it Atracci\'on del Talento Cient\'ifico en Salamanca} programme, from Project PID2021-122938NB-I00 funded by MCIN/AEI/
10.13039/501100011033 and by “ERDF A way of making Europe”,
and {\it Ayudas del Programa XIII} by USAL. DF acknowledges support from the programme {\it Ayudas para Financiar la Contrataci\'on Predoctoral de Personal Investigador (ORDEN EDU/601/2020)} funded by Junta de Castilla y Le\'on and European Social Fund. 

\clearpage
\bibliography{bibio}

\begin{thebibliography}{52}%
\makeatletter
\providecommand \@ifxundefined [1]{%
 \@ifx{#1\undefined}
}%
\providecommand \@ifnum [1]{%
 \ifnum #1\expandafter \@firstoftwo
 \else \expandafter \@secondoftwo
 \fi
}%
\providecommand \@ifx [1]{%
 \ifx #1\expandafter \@firstoftwo
 \else \expandafter \@secondoftwo
 \fi
}%
\providecommand \natexlab [1]{#1}%
\providecommand \enquote  [1]{``#1''}%
\providecommand \bibnamefont  [1]{#1}%
\providecommand \bibfnamefont [1]{#1}%
\providecommand \citenamefont [1]{#1}%
\providecommand \href@noop [0]{\@secondoftwo}%
\providecommand \href [0]{\begingroup \@sanitize@url \@href}%
\providecommand \@href[1]{\@@startlink{#1}\@@href}%
\providecommand \@@href[1]{\endgroup#1\@@endlink}%
\providecommand \@sanitize@url [0]{\catcode `\\12\catcode `\$12\catcode
  `\&12\catcode `\#12\catcode `\^12\catcode `\_12\catcode `\%12\relax}%
\providecommand \@@startlink[1]{}%
\providecommand \@@endlink[0]{}%
\providecommand \url  [0]{\begingroup\@sanitize@url \@url }%
\providecommand \@url [1]{\endgroup\@href {#1}{\urlprefix }}%
\providecommand \urlprefix  [0]{URL }%
\providecommand \Eprint [0]{\href }%
\providecommand \doibase [0]{http://dx.doi.org/}%
\providecommand \selectlanguage [0]{\@gobble}%
\providecommand \bibinfo  [0]{\@secondoftwo}%
\providecommand \bibfield  [0]{\@secondoftwo}%
\providecommand \translation [1]{[#1]}%
\providecommand \BibitemOpen [0]{}%
\providecommand \bibitemStop [0]{}%
\providecommand \bibitemNoStop [0]{.\EOS\space}%
\providecommand \EOS [0]{\spacefactor3000\relax}%
\providecommand \BibitemShut  [1]{\csname bibitem#1\endcsname}%
\let\auto@bib@innerbib\@empty
\bibitem [{\citenamefont {Aghanim}\ \emph
  {et~al.}(2020{\natexlab{a}})\citenamefont {Aghanim} \emph
  {et~al.}}]{Aghanim:2018eyx}%
  \BibitemOpen
  \bibfield  {author} {\bibinfo {author} {\bibfnamefont {N.}~\bibnamefont
  {Aghanim}} \emph {et~al.} (\bibinfo {collaboration} {Planck}),\ }\href
  {\doibase 10.1051/0004-6361/201833910} {\bibfield  {journal} {\bibinfo
  {journal} {Astron. Astrophys.}\ }\textbf {\bibinfo {volume} {641}},\ \bibinfo
  {pages} {A6} (\bibinfo {year} {2020}{\natexlab{a}})},\ \Eprint
  {http://arxiv.org/abs/1807.06209} {arXiv:1807.06209 [astro-ph.CO]}
  \BibitemShut {NoStop}%
\bibitem [{\citenamefont {Riess}\ \emph {et~al.}(1998)\citenamefont {Riess}
  \emph {et~al.}}]{SupernovaSearchTeam:1998fmf}%
  \BibitemOpen
  \bibfield  {author} {\bibinfo {author} {\bibfnamefont {A.~G.}\ \bibnamefont
  {Riess}} \emph {et~al.} (\bibinfo {collaboration} {Supernova Search Team}),\
  }\href {\doibase 10.1086/300499} {\bibfield  {journal} {\bibinfo  {journal}
  {Astron. J.}\ }\textbf {\bibinfo {volume} {116}},\ \bibinfo {pages} {1009}
  (\bibinfo {year} {1998})},\ \Eprint {http://arxiv.org/abs/astro-ph/9805201}
  {arXiv:astro-ph/9805201} \BibitemShut {NoStop}%
\bibitem [{\citenamefont {Perlmutter}\ \emph {et~al.}(1999)\citenamefont
  {Perlmutter} \emph {et~al.}}]{SupernovaCosmologyProject:1998vns}%
  \BibitemOpen
  \bibfield  {author} {\bibinfo {author} {\bibfnamefont {S.}~\bibnamefont
  {Perlmutter}} \emph {et~al.} (\bibinfo {collaboration} {Supernova Cosmology
  Project}),\ }\href {\doibase 10.1086/307221} {\bibfield  {journal} {\bibinfo
  {journal} {Astrophys. J.}\ }\textbf {\bibinfo {volume} {517}},\ \bibinfo
  {pages} {565} (\bibinfo {year} {1999})},\ \Eprint
  {http://arxiv.org/abs/astro-ph/9812133} {arXiv:astro-ph/9812133} \BibitemShut
  {NoStop}%
\bibitem [{\citenamefont {Eisenstein}\ \emph {et~al.}(2005)\citenamefont
  {Eisenstein} \emph {et~al.}}]{SDSS:2005xqv}%
  \BibitemOpen
  \bibfield  {author} {\bibinfo {author} {\bibfnamefont {D.~J.}\ \bibnamefont
  {Eisenstein}} \emph {et~al.} (\bibinfo {collaboration} {SDSS}),\ }\href
  {\doibase 10.1086/466512} {\bibfield  {journal} {\bibinfo  {journal}
  {Astrophys. J.}\ }\textbf {\bibinfo {volume} {633}},\ \bibinfo {pages} {560}
  (\bibinfo {year} {2005})},\ \Eprint {http://arxiv.org/abs/astro-ph/0501171}
  {arXiv:astro-ph/0501171} \BibitemShut {NoStop}%
\bibitem [{\citenamefont {Ross}\ \emph {et~al.}(2015)\citenamefont {Ross},
  \citenamefont {Samushia}, \citenamefont {Howlett}, \citenamefont {Percival},
  \citenamefont {Burden},\ and\ \citenamefont {Manera}}]{Ross:2014qpa}%
  \BibitemOpen
  \bibfield  {author} {\bibinfo {author} {\bibfnamefont {A.~J.}\ \bibnamefont
  {Ross}}, \bibinfo {author} {\bibfnamefont {L.}~\bibnamefont {Samushia}},
  \bibinfo {author} {\bibfnamefont {C.}~\bibnamefont {Howlett}}, \bibinfo
  {author} {\bibfnamefont {W.~J.}\ \bibnamefont {Percival}}, \bibinfo {author}
  {\bibfnamefont {A.}~\bibnamefont {Burden}}, \ and\ \bibinfo {author}
  {\bibfnamefont {M.}~\bibnamefont {Manera}},\ }\href {\doibase
  10.1093/mnras/stv154} {\bibfield  {journal} {\bibinfo  {journal} {Mon. Not.
  Roy. Astron. Soc.}\ }\textbf {\bibinfo {volume} {449}},\ \bibinfo {pages}
  {835} (\bibinfo {year} {2015})},\ \Eprint {http://arxiv.org/abs/1409.3242}
  {arXiv:1409.3242 [astro-ph.CO]} \BibitemShut {NoStop}%
\bibitem [{\citenamefont {Alam}\ \emph {et~al.}(2017)\citenamefont {Alam} \emph
  {et~al.}}]{BOSS:2016wmc}%
  \BibitemOpen
  \bibfield  {author} {\bibinfo {author} {\bibfnamefont {S.}~\bibnamefont
  {Alam}} \emph {et~al.} (\bibinfo {collaboration} {BOSS}),\ }\href {\doibase
  10.1093/mnras/stx721} {\bibfield  {journal} {\bibinfo  {journal} {Mon. Not.
  Roy. Astron. Soc.}\ }\textbf {\bibinfo {volume} {470}},\ \bibinfo {pages}
  {2617} (\bibinfo {year} {2017})},\ \Eprint {http://arxiv.org/abs/1607.03155}
  {arXiv:1607.03155 [astro-ph.CO]} \BibitemShut {NoStop}%
\bibitem [{\citenamefont {Hildebrandt}\ \emph {et~al.}(2017)\citenamefont
  {Hildebrandt} \emph {et~al.}}]{Hildebrandt:2016iqg}%
  \BibitemOpen
  \bibfield  {author} {\bibinfo {author} {\bibfnamefont {H.}~\bibnamefont
  {Hildebrandt}} \emph {et~al.},\ }\href {\doibase 10.1093/mnras/stw2805}
  {\bibfield  {journal} {\bibinfo  {journal} {Mon. Not. Roy. Astron. Soc.}\
  }\textbf {\bibinfo {volume} {465}},\ \bibinfo {pages} {1454} (\bibinfo {year}
  {2017})},\ \Eprint {http://arxiv.org/abs/1606.05338} {arXiv:1606.05338
  [astro-ph.CO]} \BibitemShut {NoStop}%
\bibitem [{\citenamefont {Abbott}\ \emph {et~al.}(2018)\citenamefont {Abbott}
  \emph {et~al.}}]{DES:2017myr}%
  \BibitemOpen
  \bibfield  {author} {\bibinfo {author} {\bibfnamefont {T.~M.~C.}\
  \bibnamefont {Abbott}} \emph {et~al.} (\bibinfo {collaboration} {DES}),\
  }\href {\doibase 10.1103/PhysRevD.98.043526} {\bibfield  {journal} {\bibinfo
  {journal} {Phys. Rev. D}\ }\textbf {\bibinfo {volume} {98}},\ \bibinfo
  {pages} {043526} (\bibinfo {year} {2018})},\ \Eprint
  {http://arxiv.org/abs/1708.01530} {arXiv:1708.01530 [astro-ph.CO]}
  \BibitemShut {NoStop}%
\bibitem [{\citenamefont {Perivolaropoulos}\ and\ \citenamefont
  {Skara}(2022)}]{Perivolaropoulos:2021jda}%
  \BibitemOpen
  \bibfield  {author} {\bibinfo {author} {\bibfnamefont {L.}~\bibnamefont
  {Perivolaropoulos}}\ and\ \bibinfo {author} {\bibfnamefont {F.}~\bibnamefont
  {Skara}},\ }\href {\doibase 10.1016/j.newar.2022.101659} {\bibfield
  {journal} {\bibinfo  {journal} {New Astron. Rev.}\ }\textbf {\bibinfo
  {volume} {95}},\ \bibinfo {pages} {101659} (\bibinfo {year} {2022})},\
  \Eprint {http://arxiv.org/abs/2105.05208} {arXiv:2105.05208 [astro-ph.CO]}
  \BibitemShut {NoStop}%
\bibitem [{\citenamefont {Di~Valentino}\ \emph
  {et~al.}(2021{\natexlab{a}})\citenamefont {Di~Valentino} \emph
  {et~al.}}]{DiValentino:2020vvd}%
  \BibitemOpen
  \bibfield  {author} {\bibinfo {author} {\bibfnamefont {E.}~\bibnamefont
  {Di~Valentino}} \emph {et~al.},\ }\href {\doibase
  10.1016/j.astropartphys.2021.102604} {\bibfield  {journal} {\bibinfo
  {journal} {Astropart. Phys.}\ }\textbf {\bibinfo {volume} {131}},\ \bibinfo
  {pages} {102604} (\bibinfo {year} {2021}{\natexlab{a}})},\ \Eprint
  {http://arxiv.org/abs/2008.11285} {arXiv:2008.11285 [astro-ph.CO]}
  \BibitemShut {NoStop}%
\bibitem [{\citenamefont {Aghanim}\ \emph
  {et~al.}(2020{\natexlab{b}})\citenamefont {Aghanim} \emph
  {et~al.}}]{Planck:2018vyg}%
  \BibitemOpen
  \bibfield  {author} {\bibinfo {author} {\bibfnamefont {N.}~\bibnamefont
  {Aghanim}} \emph {et~al.} (\bibinfo {collaboration} {Planck}),\ }\href
  {\doibase 10.1051/0004-6361/201833910} {\bibfield  {journal} {\bibinfo
  {journal} {Astron. Astrophys.}\ }\textbf {\bibinfo {volume} {641}},\ \bibinfo
  {pages} {A6} (\bibinfo {year} {2020}{\natexlab{b}})},\ \bibinfo {note}
  {[Erratum: Astron.Astrophys. 652, C4 (2021)]},\ \Eprint
  {http://arxiv.org/abs/1807.06209} {arXiv:1807.06209 [astro-ph.CO]}
  \BibitemShut {NoStop}%
\bibitem [{\citenamefont {Abbott}\ \emph {et~al.}(2020)\citenamefont {Abbott}
  \emph {et~al.}}]{DES:2020ahh}%
  \BibitemOpen
  \bibfield  {author} {\bibinfo {author} {\bibfnamefont {T.~M.~C.}\
  \bibnamefont {Abbott}} \emph {et~al.} (\bibinfo {collaboration} {DES}),\
  }\href {\doibase 10.1103/PhysRevD.102.023509} {\bibfield  {journal} {\bibinfo
   {journal} {Phys. Rev. D}\ }\textbf {\bibinfo {volume} {102}},\ \bibinfo
  {pages} {023509} (\bibinfo {year} {2020})},\ \Eprint
  {http://arxiv.org/abs/2002.11124} {arXiv:2002.11124 [astro-ph.CO]}
  \BibitemShut {NoStop}%
\bibitem [{\citenamefont {Ade}\ \emph {et~al.}(2016)\citenamefont {Ade} \emph
  {et~al.}}]{Planck:2015lwi}%
  \BibitemOpen
  \bibfield  {author} {\bibinfo {author} {\bibfnamefont {P.~A.~R.}\
  \bibnamefont {Ade}} \emph {et~al.} (\bibinfo {collaboration} {Planck}),\
  }\href {\doibase 10.1051/0004-6361/201525833} {\bibfield  {journal} {\bibinfo
   {journal} {Astron. Astrophys.}\ }\textbf {\bibinfo {volume} {594}},\
  \bibinfo {pages} {A24} (\bibinfo {year} {2016})},\ \Eprint
  {http://arxiv.org/abs/1502.01597} {arXiv:1502.01597 [astro-ph.CO]}
  \BibitemShut {NoStop}%
\bibitem [{\citenamefont {Wang}\ \emph {et~al.}(2024)\citenamefont {Wang},
  \citenamefont {Abdalla}, \citenamefont {Atrio-Barandela},\ and\ \citenamefont
  {Pav\'on}}]{Wang:2024vmw}%
  \BibitemOpen
  \bibfield  {author} {\bibinfo {author} {\bibfnamefont {B.}~\bibnamefont
  {Wang}}, \bibinfo {author} {\bibfnamefont {E.}~\bibnamefont {Abdalla}},
  \bibinfo {author} {\bibfnamefont {F.}~\bibnamefont {Atrio-Barandela}}, \ and\
  \bibinfo {author} {\bibfnamefont {D.}~\bibnamefont {Pav\'on}},\ }\href
  {\doibase 10.1088/1361-6633/ad2527} {\bibfield  {journal} {\bibinfo
  {journal} {Rept. Prog. Phys.}\ }\textbf {\bibinfo {volume} {87}},\ \bibinfo
  {pages} {036901} (\bibinfo {year} {2024})},\ \Eprint
  {http://arxiv.org/abs/2402.00819} {arXiv:2402.00819 [astro-ph.CO]}
  \BibitemShut {NoStop}%
\bibitem [{\citenamefont {Di~Valentino}\ \emph
  {et~al.}(2021{\natexlab{b}})\citenamefont {Di~Valentino}, \citenamefont
  {Mena}, \citenamefont {Pan}, \citenamefont {Visinelli}, \citenamefont {Yang},
  \citenamefont {Melchiorri}, \citenamefont {Mota}, \citenamefont {Riess},\
  and\ \citenamefont {Silk}}]{DiValentino:2021izs}%
  \BibitemOpen
  \bibfield  {author} {\bibinfo {author} {\bibfnamefont {E.}~\bibnamefont
  {Di~Valentino}}, \bibinfo {author} {\bibfnamefont {O.}~\bibnamefont {Mena}},
  \bibinfo {author} {\bibfnamefont {S.}~\bibnamefont {Pan}}, \bibinfo {author}
  {\bibfnamefont {L.}~\bibnamefont {Visinelli}}, \bibinfo {author}
  {\bibfnamefont {W.}~\bibnamefont {Yang}}, \bibinfo {author} {\bibfnamefont
  {A.}~\bibnamefont {Melchiorri}}, \bibinfo {author} {\bibfnamefont {D.~F.}\
  \bibnamefont {Mota}}, \bibinfo {author} {\bibfnamefont {A.~G.}\ \bibnamefont
  {Riess}}, \ and\ \bibinfo {author} {\bibfnamefont {J.}~\bibnamefont {Silk}},\
  }\href {\doibase 10.1088/1361-6382/ac086d} {\bibfield  {journal} {\bibinfo
  {journal} {Class. Quant. Grav.}\ }\textbf {\bibinfo {volume} {38}},\ \bibinfo
  {pages} {153001} (\bibinfo {year} {2021}{\natexlab{b}})},\ \Eprint
  {http://arxiv.org/abs/2103.01183} {arXiv:2103.01183 [astro-ph.CO]}
  \BibitemShut {NoStop}%
\bibitem [{\citenamefont {Di~Valentino}\ \emph
  {et~al.}(2021{\natexlab{c}})\citenamefont {Di~Valentino} \emph
  {et~al.}}]{DiValentino:2020zio}%
  \BibitemOpen
  \bibfield  {author} {\bibinfo {author} {\bibfnamefont {E.}~\bibnamefont
  {Di~Valentino}} \emph {et~al.},\ }\href {\doibase
  10.1016/j.astropartphys.2021.102605} {\bibfield  {journal} {\bibinfo
  {journal} {Astropart. Phys.}\ }\textbf {\bibinfo {volume} {131}},\ \bibinfo
  {pages} {102605} (\bibinfo {year} {2021}{\natexlab{c}})},\ \Eprint
  {http://arxiv.org/abs/2008.11284} {arXiv:2008.11284 [astro-ph.CO]}
  \BibitemShut {NoStop}%
\bibitem [{\citenamefont {Simpson}(2010)}]{Simpson:2010vh}%
  \BibitemOpen
  \bibfield  {author} {\bibinfo {author} {\bibfnamefont {F.}~\bibnamefont
  {Simpson}},\ }\href {\doibase 10.1103/PhysRevD.82.083505} {\bibfield
  {journal} {\bibinfo  {journal} {Phys. Rev. D}\ }\textbf {\bibinfo {volume}
  {82}},\ \bibinfo {pages} {083505} (\bibinfo {year} {2010})},\ \Eprint
  {http://arxiv.org/abs/1007.1034} {arXiv:1007.1034 [astro-ph.CO]} \BibitemShut
  {NoStop}%
\bibitem [{\citenamefont {Pourtsidou}\ and\ \citenamefont
  {Tram}(2016)}]{Pourtsidou:2016ico}%
  \BibitemOpen
  \bibfield  {author} {\bibinfo {author} {\bibfnamefont {A.}~\bibnamefont
  {Pourtsidou}}\ and\ \bibinfo {author} {\bibfnamefont {T.}~\bibnamefont
  {Tram}},\ }\href {\doibase 10.1103/PhysRevD.94.043518} {\bibfield  {journal}
  {\bibinfo  {journal} {Phys. Rev. D}\ }\textbf {\bibinfo {volume} {94}},\
  \bibinfo {pages} {043518} (\bibinfo {year} {2016})},\ \Eprint
  {http://arxiv.org/abs/1604.04222} {arXiv:1604.04222 [astro-ph.CO]}
  \BibitemShut {NoStop}%
\bibitem [{\citenamefont {Asghari}\ \emph {et~al.}(2019)\citenamefont
  {Asghari}, \citenamefont {Beltr\'an~Jim\'enez}, \citenamefont {Khosravi},\
  and\ \citenamefont {Mota}}]{Asghari:2019qld}%
  \BibitemOpen
  \bibfield  {author} {\bibinfo {author} {\bibfnamefont {M.}~\bibnamefont
  {Asghari}}, \bibinfo {author} {\bibfnamefont {J.}~\bibnamefont
  {Beltr\'an~Jim\'enez}}, \bibinfo {author} {\bibfnamefont {S.}~\bibnamefont
  {Khosravi}}, \ and\ \bibinfo {author} {\bibfnamefont {D.~F.}\ \bibnamefont
  {Mota}},\ }\href {\doibase 10.1088/1475-7516/2019/04/042} {\bibfield
  {journal} {\bibinfo  {journal} {JCAP}\ }\textbf {\bibinfo {volume} {04}},\
  \bibinfo {pages} {042} (\bibinfo {year} {2019})},\ \Eprint
  {http://arxiv.org/abs/1902.05532} {arXiv:1902.05532 [astro-ph.CO]}
  \BibitemShut {NoStop}%
\bibitem [{\citenamefont {Linton}\ \emph {et~al.}(2022)\citenamefont {Linton},
  \citenamefont {Crittenden},\ and\ \citenamefont
  {Pourtsidou}}]{Linton:2021cgd}%
  \BibitemOpen
  \bibfield  {author} {\bibinfo {author} {\bibfnamefont {M.~S.}\ \bibnamefont
  {Linton}}, \bibinfo {author} {\bibfnamefont {R.}~\bibnamefont {Crittenden}},
  \ and\ \bibinfo {author} {\bibfnamefont {A.}~\bibnamefont {Pourtsidou}},\
  }\href {\doibase 10.1088/1475-7516/2022/08/075} {\bibfield  {journal}
  {\bibinfo  {journal} {JCAP}\ }\textbf {\bibinfo {volume} {08}},\ \bibinfo
  {pages} {075} (\bibinfo {year} {2022})},\ \Eprint
  {http://arxiv.org/abs/2107.03235} {arXiv:2107.03235 [astro-ph.CO]}
  \BibitemShut {NoStop}%
\bibitem [{\citenamefont {Chamings}\ \emph {et~al.}(2020)\citenamefont
  {Chamings}, \citenamefont {Avgoustidis}, \citenamefont {Copeland},
  \citenamefont {Green},\ and\ \citenamefont {Pourtsidou}}]{Chamings:2019kcl}%
  \BibitemOpen
  \bibfield  {author} {\bibinfo {author} {\bibfnamefont {F.~N.}\ \bibnamefont
  {Chamings}}, \bibinfo {author} {\bibfnamefont {A.}~\bibnamefont
  {Avgoustidis}}, \bibinfo {author} {\bibfnamefont {E.~J.}\ \bibnamefont
  {Copeland}}, \bibinfo {author} {\bibfnamefont {A.~M.}\ \bibnamefont {Green}},
  \ and\ \bibinfo {author} {\bibfnamefont {A.}~\bibnamefont {Pourtsidou}},\
  }\href {\doibase 10.1103/PhysRevD.101.043531} {\bibfield  {journal} {\bibinfo
   {journal} {Phys. Rev. D}\ }\textbf {\bibinfo {volume} {101}},\ \bibinfo
  {pages} {043531} (\bibinfo {year} {2020})},\ \Eprint
  {http://arxiv.org/abs/1912.09858} {arXiv:1912.09858 [astro-ph.CO]}
  \BibitemShut {NoStop}%
\bibitem [{\citenamefont {Beltr\'an~Jim\'enez}\ \emph
  {et~al.}(2021)\citenamefont {Beltr\'an~Jim\'enez}, \citenamefont {Bettoni},
  \citenamefont {Figueruelo}, \citenamefont {Teppa~Pannia},\ and\ \citenamefont
  {Tsujikawa}}]{BeltranJimenez:2021wbq}%
  \BibitemOpen
  \bibfield  {author} {\bibinfo {author} {\bibfnamefont {J.}~\bibnamefont
  {Beltr\'an~Jim\'enez}}, \bibinfo {author} {\bibfnamefont {D.}~\bibnamefont
  {Bettoni}}, \bibinfo {author} {\bibfnamefont {D.}~\bibnamefont {Figueruelo}},
  \bibinfo {author} {\bibfnamefont {F.~A.}\ \bibnamefont {Teppa~Pannia}}, \
  and\ \bibinfo {author} {\bibfnamefont {S.}~\bibnamefont {Tsujikawa}},\ }\href
  {\doibase 10.1103/PhysRevD.104.103503} {\bibfield  {journal} {\bibinfo
  {journal} {Phys. Rev. D}\ }\textbf {\bibinfo {volume} {104}},\ \bibinfo
  {pages} {103503} (\bibinfo {year} {2021})},\ \Eprint
  {http://arxiv.org/abs/2106.11222} {arXiv:2106.11222 [astro-ph.CO]}
  \BibitemShut {NoStop}%
\bibitem [{\citenamefont {Kumar}\ and\ \citenamefont
  {Nunes}(2017)}]{Kumar:2017bpv}%
  \BibitemOpen
  \bibfield  {author} {\bibinfo {author} {\bibfnamefont {S.}~\bibnamefont
  {Kumar}}\ and\ \bibinfo {author} {\bibfnamefont {R.~C.}\ \bibnamefont
  {Nunes}},\ }\href {\doibase 10.1140/epjc/s10052-017-5334-3} {\bibfield
  {journal} {\bibinfo  {journal} {Eur. Phys. J. C}\ }\textbf {\bibinfo {volume}
  {77}},\ \bibinfo {pages} {734} (\bibinfo {year} {2017})},\ \Eprint
  {http://arxiv.org/abs/1709.02384} {arXiv:1709.02384 [astro-ph.CO]}
  \BibitemShut {NoStop}%
\bibitem [{\citenamefont {Skordis}\ \emph {et~al.}(2015)\citenamefont
  {Skordis}, \citenamefont {Pourtsidou},\ and\ \citenamefont
  {Copeland}}]{Skordis:2015yra}%
  \BibitemOpen
  \bibfield  {author} {\bibinfo {author} {\bibfnamefont {C.}~\bibnamefont
  {Skordis}}, \bibinfo {author} {\bibfnamefont {A.}~\bibnamefont {Pourtsidou}},
  \ and\ \bibinfo {author} {\bibfnamefont {E.~J.}\ \bibnamefont {Copeland}},\
  }\href {\doibase 10.1103/PhysRevD.91.083537} {\bibfield  {journal} {\bibinfo
  {journal} {Phys. Rev. D}\ }\textbf {\bibinfo {volume} {91}},\ \bibinfo
  {pages} {083537} (\bibinfo {year} {2015})},\ \Eprint
  {http://arxiv.org/abs/1502.07297} {arXiv:1502.07297 [astro-ph.CO]}
  \BibitemShut {NoStop}%
\bibitem [{\citenamefont {Baldi}\ and\ \citenamefont
  {Simpson}(2017)}]{Baldi:2016zom}%
  \BibitemOpen
  \bibfield  {author} {\bibinfo {author} {\bibfnamefont {M.}~\bibnamefont
  {Baldi}}\ and\ \bibinfo {author} {\bibfnamefont {F.}~\bibnamefont
  {Simpson}},\ }\href {\doibase 10.1093/mnras/stw2702} {\bibfield  {journal}
  {\bibinfo  {journal} {Mon. Not. Roy. Astron. Soc.}\ }\textbf {\bibinfo
  {volume} {465}},\ \bibinfo {pages} {653} (\bibinfo {year} {2017})},\ \Eprint
  {http://arxiv.org/abs/1605.05623} {arXiv:1605.05623 [astro-ph.CO]}
  \BibitemShut {NoStop}%
\bibitem [{\citenamefont {Beltr\'an~Jim\'enez}\ \emph
  {et~al.}(2023)\citenamefont {Beltr\'an~Jim\'enez}, \citenamefont {Di~Dio},\
  and\ \citenamefont {Figueruelo}}]{BeltranJimenez:2022irm}%
  \BibitemOpen
  \bibfield  {author} {\bibinfo {author} {\bibfnamefont {J.}~\bibnamefont
  {Beltr\'an~Jim\'enez}}, \bibinfo {author} {\bibfnamefont {E.}~\bibnamefont
  {Di~Dio}}, \ and\ \bibinfo {author} {\bibfnamefont {D.}~\bibnamefont
  {Figueruelo}},\ }\href {\doibase 10.1088/1475-7516/2023/11/088} {\bibfield
  {journal} {\bibinfo  {journal} {JCAP}\ }\textbf {\bibinfo {volume} {11}},\
  \bibinfo {pages} {088} (\bibinfo {year} {2023})},\ \Eprint
  {http://arxiv.org/abs/2212.08617} {arXiv:2212.08617 [astro-ph.CO]}
  \BibitemShut {NoStop}%
\bibitem [{\citenamefont {Lesgourgues}\ and\ \citenamefont
  {Pastor}(2006)}]{Lesgourgues:2006nd}%
  \BibitemOpen
  \bibfield  {author} {\bibinfo {author} {\bibfnamefont {J.}~\bibnamefont
  {Lesgourgues}}\ and\ \bibinfo {author} {\bibfnamefont {S.}~\bibnamefont
  {Pastor}},\ }\href {\doibase 10.1016/j.physrep.2006.04.001} {\bibfield
  {journal} {\bibinfo  {journal} {Phys. Rept.}\ }\textbf {\bibinfo {volume}
  {429}},\ \bibinfo {pages} {307} (\bibinfo {year} {2006})},\ \Eprint
  {http://arxiv.org/abs/astro-ph/0603494} {arXiv:astro-ph/0603494} \BibitemShut
  {NoStop}%
\bibitem [{\citenamefont {Hu}\ \emph {et~al.}(2015)\citenamefont {Hu},
  \citenamefont {Raveri}, \citenamefont {Silvestri},\ and\ \citenamefont
  {Frusciante}}]{Hu:2014sea}%
  \BibitemOpen
  \bibfield  {author} {\bibinfo {author} {\bibfnamefont {B.}~\bibnamefont
  {Hu}}, \bibinfo {author} {\bibfnamefont {M.}~\bibnamefont {Raveri}}, \bibinfo
  {author} {\bibfnamefont {A.}~\bibnamefont {Silvestri}}, \ and\ \bibinfo
  {author} {\bibfnamefont {N.}~\bibnamefont {Frusciante}},\ }\href {\doibase
  10.1103/PhysRevD.91.063524} {\bibfield  {journal} {\bibinfo  {journal} {Phys.
  Rev. D}\ }\textbf {\bibinfo {volume} {91}},\ \bibinfo {pages} {063524}
  (\bibinfo {year} {2015})},\ \Eprint {http://arxiv.org/abs/1410.5807}
  {arXiv:1410.5807 [astro-ph.CO]} \BibitemShut {NoStop}%
\bibitem [{\citenamefont {Figueruelo}\ \emph {et~al.}(2021)\citenamefont
  {Figueruelo} \emph {et~al.}}]{Figueruelo:2021elm}%
  \BibitemOpen
  \bibfield  {author} {\bibinfo {author} {\bibfnamefont {D.}~\bibnamefont
  {Figueruelo}} \emph {et~al.},\ }\href {\doibase
  10.1088/1475-7516/2021/07/022} {\bibfield  {journal} {\bibinfo  {journal}
  {JCAP}\ }\textbf {\bibinfo {volume} {07}},\ \bibinfo {pages} {022} (\bibinfo
  {year} {2021})},\ \Eprint {http://arxiv.org/abs/2103.01571} {arXiv:2103.01571
  [astro-ph.CO]} \BibitemShut {NoStop}%
\bibitem [{\citenamefont {Poulin}\ \emph {et~al.}(2023)\citenamefont {Poulin},
  \citenamefont {Bernal}, \citenamefont {Kovetz},\ and\ \citenamefont
  {Kamionkowski}}]{Poulin:2022sgp}%
  \BibitemOpen
  \bibfield  {author} {\bibinfo {author} {\bibfnamefont {V.}~\bibnamefont
  {Poulin}}, \bibinfo {author} {\bibfnamefont {J.~L.}\ \bibnamefont {Bernal}},
  \bibinfo {author} {\bibfnamefont {E.~D.}\ \bibnamefont {Kovetz}}, \ and\
  \bibinfo {author} {\bibfnamefont {M.}~\bibnamefont {Kamionkowski}},\ }\href
  {\doibase 10.1103/PhysRevD.107.123538} {\bibfield  {journal} {\bibinfo
  {journal} {Phys. Rev. D}\ }\textbf {\bibinfo {volume} {107}},\ \bibinfo
  {pages} {123538} (\bibinfo {year} {2023})},\ \Eprint
  {http://arxiv.org/abs/2209.06217} {arXiv:2209.06217 [astro-ph.CO]}
  \BibitemShut {NoStop}%
\bibitem [{\citenamefont {Cardona}\ and\ \citenamefont
  {Figueruelo}(2022)}]{Cardona:2022mdq}%
  \BibitemOpen
  \bibfield  {author} {\bibinfo {author} {\bibfnamefont {W.}~\bibnamefont
  {Cardona}}\ and\ \bibinfo {author} {\bibfnamefont {D.}~\bibnamefont
  {Figueruelo}},\ }\href {\doibase 10.1088/1475-7516/2022/12/010} {\bibfield
  {journal} {\bibinfo  {journal} {JCAP}\ }\textbf {\bibinfo {volume} {12}},\
  \bibinfo {pages} {010} (\bibinfo {year} {2022})},\ \Eprint
  {http://arxiv.org/abs/2209.12583} {arXiv:2209.12583 [astro-ph.CO]}
  \BibitemShut {NoStop}%
\bibitem [{\citenamefont {Beltr\'an~Jim\'enez}\ \emph
  {et~al.}(2020)\citenamefont {Beltr\'an~Jim\'enez}, \citenamefont {Bettoni},
  \citenamefont {Figueruelo},\ and\ \citenamefont
  {Teppa~Pannia}}]{BeltranJimenez:2020iyx}%
  \BibitemOpen
  \bibfield  {author} {\bibinfo {author} {\bibfnamefont {J.}~\bibnamefont
  {Beltr\'an~Jim\'enez}}, \bibinfo {author} {\bibfnamefont {D.}~\bibnamefont
  {Bettoni}}, \bibinfo {author} {\bibfnamefont {D.}~\bibnamefont {Figueruelo}},
  \ and\ \bibinfo {author} {\bibfnamefont {F.~A.}\ \bibnamefont
  {Teppa~Pannia}},\ }\href {\doibase 10.1088/1475-7516/2020/08/020} {\bibfield
  {journal} {\bibinfo  {journal} {JCAP}\ }\textbf {\bibinfo {volume} {08}},\
  \bibinfo {pages} {020} (\bibinfo {year} {2020})},\ \Eprint
  {http://arxiv.org/abs/2004.14661} {arXiv:2004.14661 [astro-ph.CO]}
  \BibitemShut {NoStop}%
\bibitem [{\citenamefont {Vagnozzi}\ \emph {et~al.}(2020)\citenamefont
  {Vagnozzi}, \citenamefont {Visinelli}, \citenamefont {Mena},\ and\
  \citenamefont {Mota}}]{Vagnozzi:2019kvw}%
  \BibitemOpen
  \bibfield  {author} {\bibinfo {author} {\bibfnamefont {S.}~\bibnamefont
  {Vagnozzi}}, \bibinfo {author} {\bibfnamefont {L.}~\bibnamefont {Visinelli}},
  \bibinfo {author} {\bibfnamefont {O.}~\bibnamefont {Mena}}, \ and\ \bibinfo
  {author} {\bibfnamefont {D.~F.}\ \bibnamefont {Mota}},\ }\href {\doibase
  10.1093/mnras/staa311} {\bibfield  {journal} {\bibinfo  {journal} {Mon. Not.
  Roy. Astron. Soc.}\ }\textbf {\bibinfo {volume} {493}},\ \bibinfo {pages}
  {1139} (\bibinfo {year} {2020})},\ \Eprint {http://arxiv.org/abs/1911.12374}
  {arXiv:1911.12374 [gr-qc]} \BibitemShut {NoStop}%
\bibitem [{\citenamefont {{Lesgourgues}}(2011)}]{2011arXiv1104.2932L}%
  \BibitemOpen
  \bibfield  {author} {\bibinfo {author} {\bibfnamefont {J.}~\bibnamefont
  {{Lesgourgues}}},\ }\href {\doibase 10.48550/arXiv.1104.2932} {\bibfield
  {journal} {\bibinfo  {journal} {arXiv e-prints}\ ,\ \bibinfo {eid}
  {arXiv:1104.2932}} (\bibinfo {year} {2011})},\ \Eprint
  {http://arxiv.org/abs/1104.2932} {arXiv:1104.2932 [astro-ph.IM]} \BibitemShut
  {NoStop}%
\bibitem [{\citenamefont {{Blas}}\ \emph {et~al.}(2011)\citenamefont {{Blas}},
  \citenamefont {{Lesgourgues}},\ and\ \citenamefont
  {{Tram}}}]{2011JCAP...07..034B}%
  \BibitemOpen
  \bibfield  {author} {\bibinfo {author} {\bibfnamefont {D.}~\bibnamefont
  {{Blas}}}, \bibinfo {author} {\bibfnamefont {J.}~\bibnamefont
  {{Lesgourgues}}}, \ and\ \bibinfo {author} {\bibfnamefont {T.}~\bibnamefont
  {{Tram}}},\ }\href {\doibase 10.1088/1475-7516/2011/07/034} {\bibfield
  {journal} {\bibinfo  {journal} {\jcap}\ }\textbf {\bibinfo {volume} {2011}},\
  \bibinfo {eid} {034} (\bibinfo {year} {2011})},\ \Eprint
  {http://arxiv.org/abs/1104.2933} {arXiv:1104.2933 [astro-ph.CO]} \BibitemShut
  {NoStop}%
\bibitem [{\citenamefont {Lewis}\ \emph {et~al.}(2000)\citenamefont {Lewis},
  \citenamefont {Challinor},\ and\ \citenamefont {Lasenby}}]{Lewis:1999bs}%
  \BibitemOpen
  \bibfield  {author} {\bibinfo {author} {\bibfnamefont {A.}~\bibnamefont
  {Lewis}}, \bibinfo {author} {\bibfnamefont {A.}~\bibnamefont {Challinor}}, \
  and\ \bibinfo {author} {\bibfnamefont {A.}~\bibnamefont {Lasenby}},\ }\href
  {\doibase 10.1086/309179} {\bibfield  {journal} {\bibinfo  {journal} {\apj}\
  }\textbf {\bibinfo {volume} {538}},\ \bibinfo {pages} {473} (\bibinfo {year}
  {2000})},\ \Eprint {http://arxiv.org/abs/astro-ph/9911177}
  {arXiv:astro-ph/9911177 [astro-ph]} \BibitemShut {NoStop}%
\bibitem [{\citenamefont {Howlett}\ \emph {et~al.}(2012)\citenamefont
  {Howlett}, \citenamefont {Lewis}, \citenamefont {Hall},\ and\ \citenamefont
  {Challinor}}]{Howlett:2012mh}%
  \BibitemOpen
  \bibfield  {author} {\bibinfo {author} {\bibfnamefont {C.}~\bibnamefont
  {Howlett}}, \bibinfo {author} {\bibfnamefont {A.}~\bibnamefont {Lewis}},
  \bibinfo {author} {\bibfnamefont {A.}~\bibnamefont {Hall}}, \ and\ \bibinfo
  {author} {\bibfnamefont {A.}~\bibnamefont {Challinor}},\ }\href {\doibase
  10.1088/1475-7516/2012/04/027} {\bibfield  {journal} {\bibinfo  {journal}
  {\jcap}\ }\textbf {\bibinfo {volume} {1204}},\ \bibinfo {pages} {027}
  (\bibinfo {year} {2012})},\ \Eprint {http://arxiv.org/abs/1201.3654}
  {arXiv:1201.3654 [astro-ph.CO]} \BibitemShut {NoStop}%
\bibitem [{\citenamefont {Brinckmann}\ and\ \citenamefont
  {Lesgourgues}(2019)}]{Brinckmann:2018cvx}%
  \BibitemOpen
  \bibfield  {author} {\bibinfo {author} {\bibfnamefont {T.}~\bibnamefont
  {Brinckmann}}\ and\ \bibinfo {author} {\bibfnamefont {J.}~\bibnamefont
  {Lesgourgues}},\ }\href {\doibase 10.1016/j.dark.2018.100260} {\bibfield
  {journal} {\bibinfo  {journal} {Phys. Dark Univ.}\ }\textbf {\bibinfo
  {volume} {24}},\ \bibinfo {pages} {100260} (\bibinfo {year} {2019})},\
  \Eprint {http://arxiv.org/abs/1804.07261} {arXiv:1804.07261 [astro-ph.CO]}
  \BibitemShut {NoStop}%
\bibitem [{\citenamefont {{Audren}}\ \emph {et~al.}(2013)\citenamefont
  {{Audren}}, \citenamefont {{Lesgourgues}}, \citenamefont {{Benabed}},\ and\
  \citenamefont {{Prunet}}}]{2013JCAP...02..001A}%
  \BibitemOpen
  \bibfield  {author} {\bibinfo {author} {\bibfnamefont {B.}~\bibnamefont
  {{Audren}}}, \bibinfo {author} {\bibfnamefont {J.}~\bibnamefont
  {{Lesgourgues}}}, \bibinfo {author} {\bibfnamefont {K.}~\bibnamefont
  {{Benabed}}}, \ and\ \bibinfo {author} {\bibfnamefont {S.}~\bibnamefont
  {{Prunet}}},\ }\href {\doibase 10.1088/1475-7516/2013/02/001} {\bibfield
  {journal} {\bibinfo  {journal} {\jcap}\ }\textbf {\bibinfo {volume} {2013}},\
  \bibinfo {eid} {001} (\bibinfo {year} {2013})},\ \Eprint
  {http://arxiv.org/abs/1210.7183} {arXiv:1210.7183 [astro-ph.CO]} \BibitemShut
  {NoStop}%
\bibitem [{\citenamefont {Aghanim}\ \emph
  {et~al.}(2020{\natexlab{c}})\citenamefont {Aghanim} \emph
  {et~al.}}]{Aghanim:2019ame}%
  \BibitemOpen
  \bibfield  {author} {\bibinfo {author} {\bibfnamefont {N.}~\bibnamefont
  {Aghanim}} \emph {et~al.} (\bibinfo {collaboration} {Planck}),\ }\href
  {\doibase 10.1051/0004-6361/201936386} {\bibfield  {journal} {\bibinfo
  {journal} {Astron. Astrophys.}\ }\textbf {\bibinfo {volume} {641}},\ \bibinfo
  {pages} {A5} (\bibinfo {year} {2020}{\natexlab{c}})},\ \Eprint
  {http://arxiv.org/abs/1907.12875} {arXiv:1907.12875 [astro-ph.CO]}
  \BibitemShut {NoStop}%
\bibitem [{\citenamefont {Brout}\ \emph {et~al.}(2022)\citenamefont {Brout}
  \emph {et~al.}}]{Brout:2022vxf}%
  \BibitemOpen
  \bibfield  {author} {\bibinfo {author} {\bibfnamefont {D.}~\bibnamefont
  {Brout}} \emph {et~al.},\ }\href {\doibase 10.3847/1538-4357/ac8e04}
  {\bibfield  {journal} {\bibinfo  {journal} {Astrophys. J.}\ }\textbf
  {\bibinfo {volume} {938}},\ \bibinfo {pages} {110} (\bibinfo {year}
  {2022})},\ \Eprint {http://arxiv.org/abs/2202.04077} {arXiv:2202.04077
  [astro-ph.CO]} \BibitemShut {NoStop}%
\bibitem [{\citenamefont {Beutler}\ \emph {et~al.}(2011)\citenamefont
  {Beutler}, \citenamefont {Blake}, \citenamefont {Colless}, \citenamefont
  {Jones}, \citenamefont {Staveley-Smith}, \citenamefont {Campbell},
  \citenamefont {Parker}, \citenamefont {Saunders},\ and\ \citenamefont
  {Watson}}]{Beutler:2011hx}%
  \BibitemOpen
  \bibfield  {author} {\bibinfo {author} {\bibfnamefont {F.}~\bibnamefont
  {Beutler}}, \bibinfo {author} {\bibfnamefont {C.}~\bibnamefont {Blake}},
  \bibinfo {author} {\bibfnamefont {M.}~\bibnamefont {Colless}}, \bibinfo
  {author} {\bibfnamefont {D.~H.}\ \bibnamefont {Jones}}, \bibinfo {author}
  {\bibfnamefont {L.}~\bibnamefont {Staveley-Smith}}, \bibinfo {author}
  {\bibfnamefont {L.}~\bibnamefont {Campbell}}, \bibinfo {author}
  {\bibfnamefont {Q.}~\bibnamefont {Parker}}, \bibinfo {author} {\bibfnamefont
  {W.}~\bibnamefont {Saunders}}, \ and\ \bibinfo {author} {\bibfnamefont
  {F.}~\bibnamefont {Watson}},\ }\href {\doibase
  10.1111/j.1365-2966.2011.19250.x} {\bibfield  {journal} {\bibinfo  {journal}
  {Mon. Not. Roy. Astron. Soc.}\ }\textbf {\bibinfo {volume} {416}},\ \bibinfo
  {pages} {3017} (\bibinfo {year} {2011})},\ \Eprint
  {http://arxiv.org/abs/1106.3366} {arXiv:1106.3366 [astro-ph.CO]} \BibitemShut
  {NoStop}%
\bibitem [{\citenamefont {Hou}\ \emph {et~al.}(2020)\citenamefont {Hou} \emph
  {et~al.}}]{Hou:2020rse}%
  \BibitemOpen
  \bibfield  {author} {\bibinfo {author} {\bibfnamefont {J.}~\bibnamefont
  {Hou}} \emph {et~al.},\ }\href {\doibase 10.1093/mnras/staa3234} {\bibfield
  {journal} {\bibinfo  {journal} {Mon. Not. Roy. Astron. Soc.}\ }\textbf
  {\bibinfo {volume} {500}},\ \bibinfo {pages} {1201} (\bibinfo {year}
  {2020})},\ \Eprint {http://arxiv.org/abs/2007.08998} {arXiv:2007.08998
  [astro-ph.CO]} \BibitemShut {NoStop}%
\bibitem [{\citenamefont {Neveux}\ \emph {et~al.}(2020)\citenamefont {Neveux}
  \emph {et~al.}}]{Neveux:2020voa}%
  \BibitemOpen
  \bibfield  {author} {\bibinfo {author} {\bibfnamefont {R.}~\bibnamefont
  {Neveux}} \emph {et~al.},\ }\href {\doibase 10.1093/mnras/staa2780}
  {\bibfield  {journal} {\bibinfo  {journal} {Mon. Not. Roy. Astron. Soc.}\
  }\textbf {\bibinfo {volume} {499}},\ \bibinfo {pages} {210} (\bibinfo {year}
  {2020})},\ \Eprint {http://arxiv.org/abs/2007.08999} {arXiv:2007.08999
  [astro-ph.CO]} \BibitemShut {NoStop}%
\bibitem [{\citenamefont {Tamone}\ \emph {et~al.}(2020)\citenamefont {Tamone}
  \emph {et~al.}}]{Tamone:2020qrl}%
  \BibitemOpen
  \bibfield  {author} {\bibinfo {author} {\bibfnamefont {A.}~\bibnamefont
  {Tamone}} \emph {et~al.},\ }\href {\doibase 10.1093/mnras/staa3050}
  {\bibfield  {journal} {\bibinfo  {journal} {Mon. Not. Roy. Astron. Soc.}\
  }\textbf {\bibinfo {volume} {499}},\ \bibinfo {pages} {5527} (\bibinfo {year}
  {2020})},\ \Eprint {http://arxiv.org/abs/2007.09009} {arXiv:2007.09009
  [astro-ph.CO]} \BibitemShut {NoStop}%
\bibitem [{\citenamefont {de~Mattia}\ \emph {et~al.}(2021)\citenamefont
  {de~Mattia} \emph {et~al.}}]{deMattia:2020fkb}%
  \BibitemOpen
  \bibfield  {author} {\bibinfo {author} {\bibfnamefont {A.}~\bibnamefont
  {de~Mattia}} \emph {et~al.},\ }\href {\doibase 10.1093/mnras/staa3891}
  {\bibfield  {journal} {\bibinfo  {journal} {Mon. Not. Roy. Astron. Soc.}\
  }\textbf {\bibinfo {volume} {501}},\ \bibinfo {pages} {5616} (\bibinfo {year}
  {2021})},\ \Eprint {http://arxiv.org/abs/2007.09008} {arXiv:2007.09008
  [astro-ph.CO]} \BibitemShut {NoStop}%
\bibitem [{\citenamefont {du~Mas~des Bourboux}\ \emph
  {et~al.}(2020)\citenamefont {du~Mas~des Bourboux} \emph
  {et~al.}}]{duMasdesBourboux:2020pck}%
  \BibitemOpen
  \bibfield  {author} {\bibinfo {author} {\bibfnamefont {H.}~\bibnamefont
  {du~Mas~des Bourboux}} \emph {et~al.},\ }\href {\doibase
  10.3847/1538-4357/abb085} {\bibfield  {journal} {\bibinfo  {journal}
  {Astrophys. J.}\ }\textbf {\bibinfo {volume} {901}},\ \bibinfo {pages} {153}
  (\bibinfo {year} {2020})},\ \Eprint {http://arxiv.org/abs/2007.08995}
  {arXiv:2007.08995 [astro-ph.CO]} \BibitemShut {NoStop}%
\bibitem [{\citenamefont {Abbott}\ \emph {et~al.}(2023)\citenamefont {Abbott}
  \emph {et~al.}}]{DES:2022urg}%
  \BibitemOpen
  \bibfield  {author} {\bibinfo {author} {\bibfnamefont {T.~M.~C.}\
  \bibnamefont {Abbott}} \emph {et~al.} (\bibinfo {collaboration} {DES, SPT}),\
  }\href {\doibase 10.1103/PhysRevD.107.023531} {\bibfield  {journal} {\bibinfo
   {journal} {Phys. Rev. D}\ }\textbf {\bibinfo {volume} {107}},\ \bibinfo
  {pages} {023531} (\bibinfo {year} {2023})},\ \Eprint
  {http://arxiv.org/abs/2206.10824} {arXiv:2206.10824 [astro-ph.CO]}
  \BibitemShut {NoStop}%
\bibitem [{\citenamefont {Heymans}\ \emph {et~al.}(2021)\citenamefont {Heymans}
  \emph {et~al.}}]{Heymans:2020gsg}%
  \BibitemOpen
  \bibfield  {author} {\bibinfo {author} {\bibfnamefont {C.}~\bibnamefont
  {Heymans}} \emph {et~al.},\ }\href {\doibase 10.1051/0004-6361/202039063}
  {\bibfield  {journal} {\bibinfo  {journal} {Astron. Astrophys.}\ }\textbf
  {\bibinfo {volume} {646}},\ \bibinfo {pages} {A140} (\bibinfo {year}
  {2021})},\ \Eprint {http://arxiv.org/abs/2007.15632} {arXiv:2007.15632
  [astro-ph.CO]} \BibitemShut {NoStop}%
\bibitem [{\citenamefont {Gelman}\ and\ \citenamefont
  {Rubin}(1992)}]{10.1214/ss/1177011136}%
  \BibitemOpen
  \bibfield  {author} {\bibinfo {author} {\bibfnamefont {A.}~\bibnamefont
  {Gelman}}\ and\ \bibinfo {author} {\bibfnamefont {D.~B.}\ \bibnamefont
  {Rubin}},\ }\href {\doibase 10.1214/ss/1177011136} {\bibfield  {journal}
  {\bibinfo  {journal} {Statistical Science}\ }\textbf {\bibinfo {volume}
  {7}},\ \bibinfo {pages} {457 } (\bibinfo {year} {1992})}\BibitemShut
  {NoStop}%
\bibitem [{\citenamefont {Heymans}\ \emph {et~al.}(2013)\citenamefont {Heymans}
  \emph {et~al.}}]{CFHTL}%
  \BibitemOpen
  \bibfield  {author} {\bibinfo {author} {\bibfnamefont {C.}~\bibnamefont
  {Heymans}} \emph {et~al.},\ }\href {\doibase 10.1093/mnras/stt601} {\bibfield
   {journal} {\bibinfo  {journal} {Mon. Not. Roy. Astron. Soc.}\ }\textbf
  {\bibinfo {volume} {432}},\ \bibinfo {pages} {2433} (\bibinfo {year}
  {2013})}\BibitemShut {NoStop}%
\bibitem [{\citenamefont {Akaike}(1974)}]{Akaike:AIC}%
  \BibitemOpen
  \bibfield  {author} {\bibinfo {author} {\bibfnamefont {H.}~\bibnamefont
  {Akaike}},\ }\href@noop {} {\bibfield  {journal} {\bibinfo  {journal} {IEEE
  Transactions on Automatic Control}\ }\textbf {\bibinfo {volume} {AC-19}},\
  \bibinfo {pages} {716} (\bibinfo {year} {1974})}\BibitemShut {NoStop}%
\end{thebibliography}%

\end{document}